\documentclass[english]{article}
\usepackage[T1]{fontenc}
\usepackage[latin9]{inputenc}
\usepackage{array}
\usepackage{float}
\usepackage{amsmath}
\usepackage{amsthm}
\usepackage{amssymb}
\usepackage{stmaryrd}
\usepackage{graphicx}

\makeatletter

\providecommand{\tabularnewline}{\\}
\numberwithin{equation}{section}
\numberwithin{figure}{section}
\theoremstyle{plain}
\newtheorem{thm}{\protect\theoremname}
  \theoremstyle{definition}
  \newtheorem{defn}[thm]{\protect\definitionname}

\makeatother

\usepackage{babel}
  \providecommand{\definitionname}{Definition}
\providecommand{\theoremname}{Theorem}

\begin{document}

\title{Deep Learning-Based BSDE Solver for Libor Market Model with Application
to Bermudan Swaption Pricing and Hedging}

\author{Haojie Wang\thanks{Corporate Model Risk, Wells Fargo Bank}, Han
Chen\thanks{Corporate Model Risk, Wells Fargo Bank}, Agus Sudjianto\thanks{Corporate Model Risk, Wells Fargo Bank},
Richard Liu\thanks{Corporate Model Risk, Wells Fargo Bank}, Qi Shen\thanks{Corporate Model Risk, Wells Fargo Bank, email contact: qi.shen@wellsfargo.com} }

\date{Revised version: September 18, 2018\\
First version: July 10, 2018}
\maketitle
\begin{abstract}
The Libor Market Model, also known as the BGM Model, is a term structure
model of interest rates. It is widely used to price interest rate
derivatives, especially Bermudan swaptions, and other exotic Libor
callable derivatives. For numerical implementation purposes, Monte
Carlo simulation is used for Libor Market Model to compute the prices
of derivatives. The PDE grid approach is not particularly feasible
due to the \textquotedblleft Curse of Dimensionality\textquotedblright .
The standard Monte Carlo method for American swaption pricing more
or less uses regression to estimate the expected value by a linear
combination of basis functions as demonstrated in the classical paper
of Longstaff and Schwartz \cite{Longstaff01}. However, the paper
\cite{Longstaff01} only provides the lower bound for American option
price. Another complexity arises from applying Monte Carlo simulation
is the computation of the sensitivities of the option, the so-called
\textquoteleft \textquoteleft Greeks\textquoteright \textquoteright ,
which are fundamental for a trader's hedging activity. Recently, an
alternative numerical method based on deep learning and backward stochastic
differential equations (BSDEs) appeared in quite a few research papers
\cite{Weinan17,Jiequn17}. For European style options the feedforward
deep neural networks (DNN) show not only feasibility but also efficiency
in obtaining both prices and numerical Greeks. The standard LMM implementation
requires dimension of five or higher in factor space even after PCA,
which cannot be solved by traditional PDE solvers, such as finite
differences or finite elements methods. In this paper, a new backward
DNN solver is proposed to price Bermudan swaptions. Our approach is
representing financial pricing problems in the form of high dimensional
stochastic optimal control problems, Forward-Backward SDEs, or equivalent
PDEs. We demonstrate that using backward DNN the high-dimensional
Bermudan swaption pricing and hedging can be solved effectively and
efficiently. A comparison between Monte Carlo simulation and the new
method for pricing vanilla interest rate options manifests the superior
performance of the new method. We then use backward DNN to calculate
prices and Greeks of Bermudan swaptions as a prelude for other Libor
callable derivatives.
\end{abstract}

\section{Introduction}

The Libor Market Model (LMM), also known as BGM Model, is a widely
used interest rate term structure model. As an extension of the Heath,
Jarrow, and Morton (HJM) \cite{HJM} model on continuous forward rates,
the LMM takes market observables as direct inputs to the model. Whereas
the HJM model describes the behavior of instantaneous forward rates
expressed with continuous compounding, LMM postulates dynamical propagation
of the forward Libor rates, which are the floating rates to index
the interest rate swap funding legs. Libor rates are simple add-on
rates for the payment period; and therefore, LMM overcomes some technical
problems, such as exploding rates, associated with the lognormal version
of the HJM model (see \cite{GararekMusiela97} for details). For vanilla
options such as cap/floor, European swaption, with appropriate numeraires
there exist even Black style formulae which can speed up the model
calibration. 

LMM is mainly implemented using Monte Carlo simulation. However, to
evaluate American and Bermudan swaptions using Monte Carlo simulation,
some additional numerical procedures have to be implemented to avoid
Monte Carlo simulation on Monte Carlo simulation and to approximate
the decision on early exercise. Some of the related works are: (1)
In 1998, Andersen proposed an approach involving a direct search for
an early exercise boundary parameterized in intrinsic value and the
values of still-alive options \cite{Andersen98}. (2) In 1997, Broadie
and Glasserman proposed an approach involving a stochastic recombining
tree \cite{Broadie97}. (3) In 1998, Longstaff and Schwartz assumed
that the value of an option, if not exercised, was a simple function
of variables observed at the exercise date in question. However, these
approaches mostly only provide pricing bounds of American options.
Another complexity is the computing of sensitivities using Monte Carlo
simulation. The sensitivities or \textquoteleft \textquoteleft Greeks\textquoteright \textquoteright{}
are fundamental quantities used by traders in hedging strategies. 

To compute derivative Greeks, the numerical PDE method is generally
the most natural alternative due to the following advantages: (1)
dependent variables of PDEs are still-alive and there is no need to
estimate them by extra numerical methods; (2) the Greeks are directly
given by the partial derivatives of the PDE solutions; (3) for various
path-dependent options only the boundary conditions need be changed.
However one well-known hard restriction in these PDE formulations
is the dimensionality in their states space. For example, for Libor6M
European swaption with expiry 10Y and tenor 20Y, there are 60 6M Libor
rates, and that requires implementation of 60 dimensional numerical
PDEs. In order to cope with this so-called \textquotedblleft Curse
of Dimensionality\textquotedblright , several traditional methods
are available in the literature, see \cite{Gerstner13,Beylkin05,Lopez-Salas},
which can be generally put into three categories. The first category
uses the Karhunen\textendash Loeve transformation to reduce the stochastic
differential equation to a lower dimensional equation. This reduction
results in a lower dimensional PDE associated to the previously reduced
SDE. The second category collects those methods which try to reduce
the dimension of the PDE itself. For example, we have mentioned dimension-wise
decomposition algorithms. The third category groups the methods which
reduce the complexity of the problem in the discretization layer,
for example, the method of sparse grids. It is inconclusive if any
of the aforementioned method can accurately solve the option pricing
in very high dimension.

Most recently, groups of computational mathematics researchers, such
as Weinan E of Princeton University, proposed new algorithms to solve
nonlinear parabolic partial differential equations (PDEs) based on
formulation of equivalent backward stochastic differential equations
(BSDEs) in high dimension (See \cite{Weinan17,Jiequn17}). Theoretically,
this approach is founded on the amazing generalized Feynman-Kac theorems
(see works by Pardoux and Peng \cite{Pardoux90}\cite{Pardoux92}).
The second step in the new approach utilizes the deep connection between
the BSDE/FBSDE theories and the stochastic optimal control theories
(See \cite{MaYong,PardouxRascanu,FlemingRishel,Pham,YongZhou,Zhang17}).
The third step borrows ideas and tools from the most recent amazing
developments in machine learning, reinforcement learning, and deep
neural networks (DNN) learning. As standardized since the early days
of machine learning the backpropagation, first proposed in a seminal
paper by Rumelhart, Hinton, and Williams\cite{Hinton86}, is naturally
embedded in neural networks. In a DNN the gradient of the value function
plays the role of a control/policy function, and the well-posed loss
function is to be optimized along the descending direction to speed
up convergence. The results are astonishing as people watch in awe
when the applications of DNN achieved great successes in facial recognition,
image reconstruction, natural speech, self-driving car, Go-game play,
and numerous others.(See \cite{Goodfellow16,Hinton12,Krizhevsky12,LeCun15})
The policy function is then approximated through a few layers of nonlinear
functional combination, convolution and other types of nesting, as
shown in the vast literature of deep learning and reinforcement learning
(See \cite{Arulkumaran} for a survey of deep learning.). In addition
the deep learning-based algorithms are highly scalable and can be
applied to BSDE, Stochastic Optimal Control, or PDE problems without
changing much of their implementation.

Our approach is also inspired by the recent developments in artificial
intelligence and deep learning researches. However, our research focus
differs from the computational mathematics world. For example, Weinan
E and co-authors have used forward DNN to solve various nonlinear
PDEs where the classical numerical methods suffer from the \textquotedblleft Curse
of Dimensionality\textquotedblright . They used forward DNN to solve
high dimensional parabolic PDEs in particular through BSDE representations
based on nonlinear Feynman-Kac formulae. In their examples of European
style derivatives pricing, their method projects the option value
forward with prescribed SDE to the terminal time and approximates
the gradient of option value function in a feedforward DNN. The loss
function is proposed to minimize the error between the prescribed
terminal value and the projected value at terminal time. The training
data are simulated sample paths of standard Brownian Motions. Thus
it belongs to the rule-based learning category. 

For comparison purposes, we recognize that the conventional forward
DNN could be used only in European style derivatives pricing. In financial
engineering, the more complex derivatives are callable instruments,
such as callable bonds, convertible bonds, American swaptions, and
especially Bermudan swaptions. Bermudan swaptions are widely used
to hedge interest rate risk by swap desks and other fixed income desks.
However, the conventional forward solver is suitable only for European
style financial derivatives pricing. In general the pricing of Bermudan
style derivatives needs to handle early exercise appropriately. A
general guideline is the famous Bellman dynamic programming principle
to make this type of optimal decision. For example, a recombining
tree or PDE grids are used in low dimensional implementation of Bermudan
swaption pricing, and the valuation process is backward starting from
the terminal time. At each exercise date the immediate exercise value
is compared to the computed holding value of the option to make an
optimal decision for exercise. 

According to our knowledge our paper is the first in the financial
engineering literature to apply backward DNN to price callable derivatives.
We have developed a new method which uses a similar DNN structure/layers
but marching backwards to project flows. The idea is quite natural
and consistent to BSDE formulation. Our approach is specifically designed
for callable derivatives pricing. When option price is projected backwards,
it is easy to make an early exercise decision of Bermudan swaption
following the same Bellman dynamic programming principle. Instead
of matching the final payoff at terminal time, our new approach learns
the parameters by minimizing the variance of the projected initial
values from different simulated paths. We thus call this new approach
the backward DNN learning. Clearly, the approach demonstrated in our
paper can readily be applied to general callable derivatives pricing
in financial engineering. Our approach is directly taking the financial
engineering problem in hand and makes the use of intrinsic connections
among three major fields of knowledge: Stochastic Optimal Control,
FBSDE theory, and PDE theories, especially parabolic and elliptic
PDEs. We can represent the financial derivative pricing problem in
any of these three formulations and choose the one that is most computationally
efficient.

The paper is organized in five sections. In Section \ref{sec:Description-of-Libor},
we introduce the basics about Libor market model, including probability
measure, pricing numeraire, and dynamic evolution. In Section \ref{sec:Deep-Learning-Implementation},
we first represent the financial derivative pricing problems in PDE
and BSDE formulations under LMM. We then discuss the deep learning-inspired
forward and backward DNN solvers. We explain the major contribution
of this paper of backward-feeding DNN algorithm in details, and its
unique advantage to solve the pricing problem for callable options.
We also compare the forward and backward DNN to highlight their differences,
advantages, and disadvantages. In Section \ref{sec:Numerical-Results},
we use the LMM setting in QuantLib\footnote{QuantLib is a free/open-source library for quantitative finance.}
to conduct some numerical experiments to show the performance of forward
and backward DNN solvers such as numerical accuracy, convergence speed,
and stability. For this purpose, we have drawn comparisons between
our backward DNN results and Monte Carlo simulation results from QuantLib
in the pricing of European swaptions. This serves as the benchmark
for our further development on Bermudan swaptions evaluated using
DNN backward solver. Concretely, we analyze the performance of backward
DNN solver in pricing Bermudan swaption by increasing the number of
exercise dates and using different spot yield curves. We conclude
our paper in Section \ref{sec:Conclusion} by pointing out our view
on future DNN research directions in financial engineering and derivatives
pricing and hedging.

\section{Description of Libor Market Model\label{sec:Description-of-Libor}}

In this section, we introduce the Libor market model (LMM) and set
up the theoretical framework for numerical implementation using deep
neural networks. We mostly follow notations in Andersen-Piterbarg
\cite{AndersenPiterbarg}. Later, our neural network pricing approach
is discussed in the context of interest rate options under the LMM
setting. Bear in mind that our methodology can be applied to other
general dynamics and financial derivative products. 

\subsection{LMM Dynamics and Pricing Measures}

Let $T>0$ and $(\Omega,\mathcal{F},P,\mathbb{\,F})$ be a complete
filtered probability space satisfying the usual conditions, where
the filtration $\mathbb{F}=\left\{ \mathcal{F}_{t}\right\} _{0\leq t\leq T}$
is the natural filtration generated by the standard Brownian motion
$\left\{ W_{t}\right\} _{0\leq t\leq T}$ (possibly high-dimensional)
and augmented by all $P-$null sets.

In the implementation of the Libor market model, it is customary to
start from the definition of the fixed tenor structure. The model
tenor structure is a set of dates, i.e. 

\begin{equation}
0=T_{0}<T_{1}<\cdots<T_{N}
\end{equation}
characterized by intervals $\tau_{i}=\tau(T_{i+1}-T_{i})$, $\forall i=0,1,\cdots,N-1$.
We then specify the vector standard Brownian Motions $W^{n}(t),\:n=1,2,...,N-1$
for all generated $\mathbb{F}=\left\{ \mathcal{F}_{t}\right\} _{0\leq t\leq T}$.
The function values are typically set to be either three or six months
corresponding to a pre-specified day-count basis indicated by the
function $\tau(x)$. Here are two examples for day-count fractionals:

\[
\tau(T_{i+1}-T_{i})=\frac{T_{i+1}-T_{i}}{365};
\]

\[
\tau(T_{i+1}-T_{i})=\frac{T_{i+1}-T_{i}}{360}.
\]

Now we recall several important definitions and propositions.
\begin{defn}
(Forward Libor Rate) Let $P(t,T)$ denote the time $t$ price of a
zero-coupon bond delivering for certain \$1 at maturity $T$, for
the tenor structure aforementioned. The forward Libor rates are defined
by:
\end{defn}

\begin{equation}
L(t;T_{n},T_{n+1})\triangleq L_{n}(t)=\frac{1}{\tau_{n}}\left(\frac{P(t,T_{n})}{P(t,T_{n+1})}-1\right),\;\;1\leq q(t)\le n\le N-1,
\end{equation}
where $q(t)$ represents the unique integer such that 

\[
T_{q(t)-1}\leq t<T_{q(t)},\;\:\forall t\leq T_{N}\,.
\]

We denote $L(t)\,\triangleq\,\left(L_{1}(t),\,L_{2}(t),\,...,L_{N-1}(t)\right)$.

$L_{n}(t)$ must be a martingale in the $T_{n+1}$- forward measure
$Q^{T_{n}+1}$ such that $L_{n}(t)$ follows SDE\footnote{In our subsequent analysis and implementation we assmue the duffusion
coefficient is a loacl volatility function $\sigma\left(t,L_{n}(t)\right)\triangleq\psi(t)L_{n}(t),$that
is, we work on a lognowmal version of LMM. }:

\begin{equation}
dL_{n}(t)=\sigma_{n}(t,\,L_{n}(t))dW^{n+1}(t)
\end{equation}
where $W^{n+1}(t)=W^{T_{n+1}}$ is a \footnote{$d\leq N-1\,$is a positive integer in general.}$d$-dimensional
Brownian motion in forward measure $Q^{T_{n+1}}$, and $\sigma_{n}\left(t,L_{n}(t)\right)$is
a row vector in $R^{d}$ . 

To simplify the notation we denote $\xi_{i}(t,L_{n}(t))\,\triangleq\bigparallel\sigma_{i}(t,L_{n}(t))\parallel\in R^{1}.$
Under the terminal measure $Q^{T_{N}}$, the process for $L_{n}(t)$
is 

\begin{equation}
\begin{gathered}dL_{n}(t)=\sigma_{n}(t,L_{n}(t))\left(-\sum_{j=n+1}^{N-1}\frac{\tau_{j}\sigma_{j}(t,L_{j}(t))}{1+\tau_{j}L_{j}(t)}dt+dW^{N}(t)\right)\\
=-\sum_{j=n+1}^{N-1}\frac{\tau_{j}\xi_{n}(t,L_{n}(t))\xi_{j}(t,L_{j}(t))\rho_{n,j}}{1+\tau_{j}L_{j}(t)}dt+\xi_{n}(t,L_{n}(t)dW_{n}^{N}(t)\\
\triangleq\:\mu_{n}(t,L_{n}(t))dt+\xi_{n}(t,L_{n}(t))dW_{n}^{N}(t)
\end{gathered}
\label{eq:Libor-SDE-Terminal-Measure}
\end{equation}
where $W_{n}^{N}(t)$ is a one-dimensional standard Brownian motion
for the forward Libor rate $L_{n}(t)$ in terminal measure $Q^{T_{N}}.$
The $\Vert\bullet\Vert$is the Euclidean-norm in $R^{d}.$ We have
assumed the constant correlation $\left\{ \rho_{ij}\right\} $for
components of $W^{N}.$ We can write explicitly for the drift term
as:

\begin{equation}
\begin{gathered}\mu_{n}(t)=-\sum_{j=n+1}^{N-1}\frac{\tau_{j}\xi_{j}(t,L_{j}(t))\xi_{n}(t,L_{n}(t))\rho_{n,j}}{1+\tau_{j}L_{j}(t)}dt\\
dW_{n}^{N}(t)dW_{j}^{N}(t)=\rho_{n,j}dt
\end{gathered}
\end{equation}
Under the spot measure $Q^{B}$, the process for $L_{n}(t)$ is

\begin{equation}
\begin{gathered}dL_{n}(t)=\sigma_{n}(t,L_{n}(t))\left(\sum_{j=q(t)}^{n}\frac{\tau_{j}\sigma_{j}(t,L_{j}(t))}{1+\tau_{j}L_{j}(t)}dt+dW^{B}(t)\right)\\
=\sum_{j=q(t)}^{n}\frac{\tau_{j}\xi_{n}(t,L_{n}(t))\xi_{j}(t,L_{j}(t))\rho_{n,j}}{1+\tau_{j}L_{j}(t)}dt+\xi_{n}(t,L_{n}(t))dW_{n}^{B}(t)\\
\triangleq\mu_{n}(t,L_{n}(t))dt+\xi_{n}(t,L_{n}(t))dW_{n}^{B}(t)
\end{gathered}
\label{eq:Libor-SDE-Spot-Measure}
\end{equation}
where $W^{B}(t)=W^{B}$ is a $d$-dimensional Brownian motion in spot
measure $Q^{B}$, and $W_{n}^{B}(t)$ is a one-dimensional Brownian
motion driving forward Libor rate $L_{n}(t)$ in spot measure $Q^{B}$.
We can write explicitly for this case the drift term as:
\begin{equation}
\begin{gathered}\mu_{n}(t)=\sum_{j=q(t)}^{n}\frac{\tau_{j}\xi_{n}(t,L_{n}(t))\xi_{j}(t,L_{j}(t))\rho_{n,j}}{1+\tau_{j}L_{j}(t)}dt\\
dW_{n}^{B}(t)dW_{j}^{B}(t)=\rho_{n,j}dt
\end{gathered}
\end{equation}
Under the spot measure, the numeraire is

\begin{equation}
B(t)=P(t,T_{q(t)})\prod_{n=0}^{q(t)-1}\left(1+\tau_{n}L_{n}(T_{n})\right)
\end{equation}
which is simple compounded discrete money market account. 

\subsection{Separable Deterministic Volatility Functinon}

In order to make the Libor Market Model Markovian, we have to specify
the vector volatility function$\sigma_{n}(t)\:n=1,2,...,N-1$, in
the following form (See Andersen-Piterbarg \cite{AndersenPiterbarg}):

\begin{equation}
\sigma_{n}(t,L_{n}(t))=\lambda_{n}(t)\varphi(L_{n}(t))
\end{equation}
where $\lambda_{n}(t)$ is a bounded row-vector of deterministic functions
and $\varphi:\;R\to R$ is a time-homogeneous local volatility function.
Some standard parameterizations of $\varphi$ are shown in the following
Table:

\begin{table}[H]
\begin{tabular}{|c|c|}
\hline 
Name & $\varphi$(x)\tabularnewline
\hline 
\hline 
Log-normal & $x$\tabularnewline
\hline 
CEV & $x^{p},\;0<p<1$\tabularnewline
\hline 
LCEV & $x\;min\left(\varepsilon^{p-1},x^{p-1}\right),\quad0<p<1,\;\varepsilon>0$\tabularnewline
\hline 
Displaced log-normal & $bx+a,\;b>0,\;a\ne0$\tabularnewline
\hline 
\end{tabular}

\caption{Specification of local volatility function}

\end{table}

\subsection{Correlation Structure}

In general, the LMM can capture non-trivial curve movements, including
not only parallel shifts, but also ``rotational steepening'' and
``humps'', which is achieved through the use of vector-valued Brownian
motion drivers with correlation. Later in the simulation, we use the
simple correlation structure such that the Libor rates $L_{i}(t),\;i=1,2,\cdots,\,N-1$
are controlled by the correlation function: $\rho_{ij}=exp(-\beta|i-j|)$.
This is a simple postulation but is sufficient for our current work
to demonstrate the effectiveness of the backward deep neural networks.
For the subsequent sections we will fix $d=N-1$, that is, each $L_{i}(t)$
is driven by one Brownian Motion $W_{i}^{N}(t)\in R^{1}$.

With this postulation we can now write the Brownian Motion $dW^{N}(t)=\varrho^{\frac{1}{2}}dW(t)$,
$\varrho=\left(\rho_{ij}\right),i,j=1,2,...,N-1.$ Here, $W(t)$ is
a standard Brownian Motion in $R^{N-1}$.

\section{Deep Neural Network Implementation\label{sec:Deep-Learning-Implementation}}

In this section, we will first state the PDE/BSDE formulation for
the European/Bermudan pricing for LMM setting, and then move on to
the neural network solver for the derived PDE/BSDE problem.

\subsection{PDE Derivation for European Option Pricing\label{subsec:PDE-Derivation-European}}

An European option is characterized by its payoff function $G$, which
determines the amount $G(L(T))$ that the option pays at option expiry
$t=T$. The arbitrage-free discounted value function of the option
relative to a numeraire $A(t)$ (discounted option price) is then
given by:

\begin{equation}
u(t,\,L(t))=A(t)E^{Q^{A}}\left(\frac{G(L(T))}{A(T)}|F_{t}\right)\label{eq:discount}
\end{equation}
where $L(t)\in R^{N-1}$ is the vector of all forward Libor rates
$L_{i}(t),\;i=1,2,\cdots,\,N-1$. Using It�\textquoteright s formula,
the stochastic differential equation for $u\left(t,\,L(t)\right)$
is given by:

\begin{equation}
du(t,L(t))=u_{t}dt+\sum_{i=1}^{N-1}u_{i}dL_{i}(t)+\frac{1}{2}\sum_{i,j=1}^{N-1}u_{ij}dL_{i}(t)dL_{j}(t)\label{eq:u-SDE}
\end{equation}
where $u_{i}=u_{L_{i}}$and $u_{ij}=u_{L_{i}L_{j}}.$ 

Substituting Eq. (\ref{eq:Libor-SDE-Terminal-Measure}) under terminal
measure $Q^{T_{N}}$ or (or similar Eq. (\ref{eq:Libor-SDE-Spot-Measure})
under spot measure $Q^{B}$) in Eq. (\ref{eq:u-SDE}), we have

\begin{equation}
\begin{gathered}du(t,L(t))=\left(u_{t}+\sum_{i=1}^{N-1}\mu_{i}(t)u_{i}+\frac{1}{2}\sum_{i,j=1}^{N-1}\xi_{i}(t)\xi_{j}(t)\,u_{ij}\right)dt\\
+\sum_{i=1}^{N-1}\xi_{i}(t)\,u_{i}dW_{i}^{N}(t)
\end{gathered}
\end{equation}
 In order to comply with the arbitrage-free condition, the process
$u(t,\,L(t))$ has to be a martingale under the measure $Q^{A}$which
corresponds to the numeraire process $A(t)$ . Thus, to satisfy this
requirement, the drift term must be equal to zero. We then obtain
the PDE as following:

\begin{equation}
u_{t}+\sum_{i=1}^{N-1}\mu_{i}(t)u_{i}+\frac{1}{2}\sum_{i,j=1}^{N-1}\rho_{i,j}\xi_{i}(t)\xi_{j}(t)\,u_{ij}=0
\end{equation}
We can also write PDE (3.4) in the matrix as following :

\begin{equation}
\begin{gathered}du(t,L(t))=u_{t}dt+\nabla u\,dL(t)+\frac{1}{2}dL(t)^{T}\left(Hess_{L}u\right)dL(t)\\
=\left(u_{t}+\nabla u\,\mu+\frac{1}{2}Tr\left(\Sigma^{T}\Sigma Hess_{L}u\right)\right)dt+\nabla u\Sigma dW(t)
\end{gathered}
\label{eq:discounted-v-sde}
\end{equation}
It follows that the PDE is:

\begin{equation}
u_{t}+\nabla u\,\mu+\frac{1}{2}Tr\left(\Sigma^{T}\Sigma Hess_{L}u\right)=0\label{eq:swaption-PDE}
\end{equation}
where $\mu$ is $\left(N-1\right)$-dimensional vector, and $\Sigma$
is the $(N-1)\times(N-1)$ matrix $\left(\xi_{i}\xi_{j}\rho_{ij}\right),\:i,j=1,2,...,N-1$.

\subsection{Equivalent BSDE for European Option Pricing\label{subsec:Equivalent-BSDE-European}}

By the basic BSDE theory the solution of the parabolic PDE (\ref{eq:swaption-PDE})
is connected to the following decoupled FBSDE

\[
dL(t)=\mu(t,L(t))dt+\Sigma(t,L(t))dW(t),
\]

\begin{equation}
\begin{cases}
dY(t)= & Z(t)\Sigma(t,L(t))dW(t),\\
Y(T)= & \frac{G(L(T))}{A(T)}\triangleq g(L(T)),
\end{cases}\label{eq:BSDE}
\end{equation}
where we define function $g(t)\triangleq\frac{G(L(t))}{A(t)}$. A
standard generalized Feynman-Kac formula states:

\begin{equation}
Y(t)=u(t,L(t)),\ \ \ \:Z(t)=\nabla_{L}u(t,L(t)).\label{relation}
\end{equation}
Notice that the gradient is a row vector. Actually the BSDE for our
problem can be obtained as a particular case of the general BSDE:
\begin{equation}
du(t,L(t))=\nabla_{L}u\,\Sigma dW(t)\label{eq:Option-BSDE}
\end{equation}
For more details on the BSDE applications in finance, we refer to
\cite{Karoui,Pardoux90}.

\subsection{From European Option to Bermudan Option \label{subsec:From-European-to-Bermudan}}

In this section,we first state the basics of Bermudan swaptions in
the Libor market model setting. We then explain how to construct PDE/BSDE
for a Bermudan swaption using results from sections \ref{subsec:PDE-Derivation-European}
and \ref{subsec:Equivalent-BSDE-European}. 

Bermudan option is a type of derivative securities with early exercises
that could take place on a discrete set of dates. It is characterized
by an adapted payout process $U(t,L(t))$, payable to the option holder
at a stopping time (an exercise date) $\tau\leq T,$ chosen by the
option holder. Let the allowed (and deterministic) set of exercise
dates larger than or equal to $t$ be denoted by $\mathcal{D}(t)$;
and suppose that we are given at time $0$ a particular exercise policy
$\tau$ taking values in $\mathcal{D}(0),$ as well as a pricing numeraire
$A(t)$ corresponding to a unique martingale measure $Q^{A}$. Let
$V^{\tau}(0)$ be the time $0$ value of a derivative security that
pays $U(\tau,L(\tau))$ at exercise time. Under some technical conditions
on $U(t,L(t)),$ we can write for the value of the derivative security
as:

\[
V^{\tau}(0)=A(0)\,E^{Q^{A}}\left(\frac{U(\tau,L(\tau))}{A(\tau)}\right),
\]

Let $\mathcal{T}(t)$ be the time $t$ set of (future) stopping times
taking values in $\mathcal{D}(t)$. In the absence of arbitrage, the
time $t$ value of a security with early exercise into $U$ is then
given by the optimal stopping problem:

\[
V(t,L(t))=\sup_{\tau\in\mathcal{T}(t)}V^{\tau}(t)=\sup_{\tau\in\mathcal{T}(t)}A(t)E^{Q^{A}}\left(\frac{U(\tau,L(\tau))}{A(\tau)}\right).
\]

Suppose we have $\mathcal{D}(0)={t_{k_{1}},t_{k_{2}},\ldots,t_{k_{p}}}$
where $t_{k_{p}}=T.$ (The unusual notation here is used for the convenience
of the discretization later in \ref{subsec:BackwardSolver}). For
$t<t_{k_{i+1}},$ define $H_{i}(t,L(t))$ as the time $t$ value of
the Bermudan option when exercise is restricted to the dates $\mathcal{D}(t_{k_{i+1}})={t_{k_{i+1}},t_{k_{i+2}},\ldots,t_{k_{p}}}$.
That is 
\[
H_{i}(t,L(t))=A(t)\cdot E_{t}^{Q^{A}}\left[\frac{V(t_{k_{i+1}},L(t_{k_{i+1}}))}{A(t_{k_{i}+1})}\right],\ \ i=1,\ldots,p-1.
\]

At time $t_{k_{i}},$ $H_{i}(t_{k_{i}},L(t_{k_{i}}))$ can be interpreted
as the hold value of the Bermudan option, that is, the value of the
Bermudan option if not exercised at time $t_{k_{i}}$. If an optimal
exercise policy is followed, clearly we must have at time $t_{k_{i}}$

\[
V(t_{k_{i}},L(t_{k_{i}}))=\max(U(t_{k_{i}},L(t_{k_{i}})),H_{i}(t_{k_{i}},L(t_{k_{i}}))),\ \ i=1,\ldots,p-1,
\]

such that for $i=1,\ldots,p-1$,

\begin{equation}
H_{i}(t,L(t))=A(t)\,E_{t}^{Q^{A}}(\max(U(t_{k_{i+1}},L(t_{k_{i}})),\,\frac{H_{i}(t_{k_{i}+1},L(t_{k_{i}+1}))}{A(t_{k_{i}+1})})).\label{eq:hold-value-iteration}
\end{equation}

Starting with the terminal condition $\frac{H_{p}(T,L(T))}{A(t)}=g\left(L(T)\right)$,
Eq.\ref{eq:hold-value-iteration} defines a useful iteration backwards
in time for the value $V(0)=H_{0}(0)$. In more details, for an $i\in{1,\ldots,p-1},$
we can price $\frac{H_{i}(t,L(t))}{A(t)}$ for $t\in[t_{k_{i-1}},t_{k_{i}}]$
similarly as $u(t,L(t))$ in Eq.\ref{eq:discount} given that $\frac{H_{i+1}(t,L(t))}{A(t)}$
is already priced in advance. Therefore, the processes discussed in
sections\ref{subsec:PDE-Derivation-European} and \ref{subsec:Equivalent-BSDE-European}
can be applied with backward iteration in time for sub-intervals $[t_{k_{i-1}},t_{k_{i}}]$.
Later we will see the essence of our backward neural network solver
is to take advantage of this backward iteration property and to parameterize
and learn/approximate all $\frac{H_{i}(t,L(t))}{A(t)},\:i=1,2...,\,p-1$
altogether in one backward discretization run instead of learning
them one by one. It is clear that this approach exactly follows the
famous Bellman Dynamic Programming principle and the finding of the
optimal exercise policy is assured.

\subsection{Deep Neural Network-Based BSDE Forward/Backward Solvers for the LMM }

In this section, we mainly cover (1) forward solvers; and (2) backward
solvers in using DNN. Forward solvers have been developed mainly by
Weinan E et al. \cite{Weinan17,Jiequn17}. The forward feed DNN is
suitable to price European style options, but it presents difficulties
to price Bermudan options as explained below:
\begin{itemize}
\item For a Bermudan style option, there are multiple exercise dates. At
any future time of an exercise date, according to the dynamic programming
principle for optimality, the continuation value of the option must
be known as well. Given a numerical scheme for pricing, forward estimation
of the continuation value could be very difficult. For example, Monte
Carlo simulation scheme is generally forward looking and one needs
special method, such as regression, or basis function approximation
to estimate the expected continuation value.
\end{itemize}
Therefore, it is generally infeasible or inefficient to use forward
solvers to price Bermudan swaptions. For neural network implementation
it encounters similar problem such as those simulation methods (See
\cite{Longstaff01,Andersen98,Broadie97}) encountered. The major contribution
of our work is to explore a new approach: Backward DNN solver. With
our limited knowledge, this paper is the first work in the literature
in the DNN application to derivatives pricing area to show that the
backward DNN is effective and efficient for pricing Bermudan options.

We provide the details of these two solvers in the next two subsections.

\subsubsection{Discretization of Libor Rate SDE and Option Price BSDE}

To derive the deep neural network-based BSDE solver, we first need
to discretize the Libor rate SDE and the option price BSDE. Before
the discretization, the time discretization is set to be:
\begin{equation}
0=t_{0}<t_{1}<\cdots<t_{m}=T_{N-1}
\end{equation}

where $m$ is the total number of grid points and the terminal time/last
grid point is $T=T_{N-1}$.

We mainly discretize Libor rate SDE in Eq. (\ref{eq:Libor-SDE-Terminal-Measure}
or \ref{eq:Libor-SDE-Spot-Measure}) under terminal measure or spot
measure for Libor rates $L_{n}(t)$: 
\begin{equation}
\begin{gathered}L_{n}(t_{i+1})\approx L{}_{n}(t_{i})+\int_{t_{i}}^{t_{i+1}}\mu_{n}(t,L(t))dt+\int_{t_{i}}^{t_{i+1}}\xi_{n}(t,L(t))dW_{n}(t)\end{gathered}
\label{eq:Libor-SDE-Discrete}
\end{equation}
for $0\le i\le N-1$. There are quite a few discretization schemes
to discretize the Libor SDEs, among which are Euler scheme and predictor-corrector
scheme. 

For the discretization of the BSDE of discounted option price, there
is a slight difference between the forward solver and the backward
solver. Their descretizations are listed as follows.
\begin{itemize}
\item \textbf{Forward solver:}
\end{itemize}
\begin{equation}
u\left(t_{i+1},L(t_{i+1})\right)\approx u(t_{i},L(t_{i}))+\nabla u\left(t_{i},L(t_{i})\right)\sigma\left(t_{i},L(t_{i})\right)\left(W\left(t_{n+1}\right)-W\left(t_{n}\right)\right)\label{eq:Option-Price-Approx-Fwd-Solver}
\end{equation}
for $0\le i\le N-1$.
\begin{itemize}
\item \textbf{Backward solver:}
\end{itemize}
\begin{equation}
u(t_{i},L(t_{i}))\approx u\left(t_{i+1},L(t_{i+1})\right)-\nabla u\left(t_{i},L(t_{i})\right)\sigma\left(t_{i},L(t_{i})\right)\left(W\left(t_{n+1}\right)-W\left(t_{n}\right)\right)\label{eq:Option-Price-Approx-Back-Solver}
\end{equation}
for $0\le i\le N-1$, which is obtained by reordering the terms in
Eq.(\ref{eq:Option-Price-Approx-Fwd-Solver}).

\subsubsection{Multilayer Feedforward Neural Network Based Algorithms}

Given this temporal discretization, the path $\left\{ L(t)_{0\le t\le T_{N-1}}\right\} $
can be easily sampled using Eq. (\ref{eq:Libor-SDE-Discrete}). Our
key step next is to approximate the function $L(t)\to\nabla u\left(t,L(t)\right)\sigma\left(t,L(t)\right)$
at each time step $t=t_{i}$ by a multilayer feed forward neural network
for $i=1,2,\ldots,N-1$
\begin{equation}
\nabla u\left(t_{i},L(t_{i})\right)\sigma\left(t_{i},L(t_{i})\right)=\left(\nabla u\,\sigma\right)\left(t_{i},L(t_{i})\right)\approx\left(\nabla u\sigma\right)\left(t_{i},L(t_{i})|\theta_{i}\right),\label{eq:NN-Gradient}
\end{equation}

where $\theta_{i}$ denotes the parameters of the neural network approximating
$L(t)\to\nabla u\left(t,L(t)\right)\sigma\left(t,L(t)\right)$ at
$t=t_{i}$. Thereafter, we have stacked all the sub-networks in Eq.
(\ref{eq:NN-Gradient}) together to form a deep neural network as
a whole, based on Eq. (\ref{eq:Option-Price-Approx-Fwd-Solver}) or
Eq. (\ref{eq:Option-Price-Approx-Back-Solver}). Specifically, this
network takes the Libor rate path $\left\{ L(t_{i})_{0\le i\le m}\right\} $
and Brownian motion path$\left\{ W(t_{i})_{0\le i\le m}\right\} $
as the input data and provides the final outputs. The forward and
backward solvers are different in approximating the target function.
For forward solver, the output is the approximated terminal payoff;
while for the backward solver the initial value and gradient is the
output. For detail, we refer to the following two subsections. 

\subsubsection{Stochastic Optimization Algorithms for Forward DNN Solver }

First we apply the forward solver methodology \cite{Weinan17,Jiequn17}
to price European option in the Libor market model setting. We basically
use Eqs. (\ref{eq:Libor-SDE-Discrete}, \ref{eq:Option-Price-Approx-Fwd-Solver})
to project Libor rates and discounted option price samples forward
from $\left(t_{0},L(t_{0}),u(t_{0},L(t_{0}))\right)$ to $\left(T_{N-1},L(T_{N-1}),u(T_{N-1},L(T_{N-1}))\right)$.
The final outputs for BSDE in Eq. (\ref{eq:Option-Price-Approx-Fwd-Solver})
are $\hat{u}(T_{N-1},L(T_{N-1}))$, as an approximation of final actual
discounted option payoff $g(L(T_{N-1}))$. The whole network flow
is the following:
\begin{itemize}
\item Input: Brownian motion path$\left\{ W(t_{i})_{0\le i\le N-1}\right\} $
and the Libor rate path $\left\{ L(t_{i})_{0\le i\le N-1}\right\} $
from Eq. (\ref{eq:Libor-SDE-Discrete}).
\item Parameters: $\theta=\left\{ \theta_{u_{0}},\theta_{\nabla_{L}u},\;\theta_{1},\;\ldots,\;\theta_{N-1}\right\} $.
The first parameter is the initial discounted option price, the second
parameters are the gradients w.r.t underlying Libor rates at the initial
point, and the remaining components of $\theta$ are the parameters
used to approximate the option price gradients w.r.t Libor rates $L(t_{0})$.
The parameters across all simulated Libor samples are identical.
\item Forward projection iterations: for $i=1,\ldots,N-1$,
\end{itemize}
\begin{equation}
\hat{u}\left(t_{i+1},L(t_{i+1})\right)=\hat{u}(t_{i},L(t_{i}))+\left(\nabla u\sigma\right)\left(t_{i},L(t_{i})|\theta_{i}\right)\left(W\left(t_{i+1}\right)-W\left(t_{i}\right)\right)\label{eq:Fwd-flow}
\end{equation}

where the initial price parameter $\hat{u}\left(t_{0},L(t_{0})\right)$
=$\theta_{u_{0}}$and the final output is $\hat{u}\left(T_{N-1},L(T_{N-1})\right)$. 

After obtaining the neural network approximated final discounted payoff
and actual discounted payoff, we need minimize the expected loss function
below:
\begin{equation}
l\left(\theta\right)=E\left[|\hat{u}\left(T_{N-1},L(T_{N-1})\right)-g(L(T_{N-1}))|^{2}\right]
\end{equation}
when $S$ samples of Monte Carlo simulation are used, $l(\theta)$
can also be re-written as: 

\begin{equation}
\begin{gathered}l\left(\theta\right)=E\left[|\hat{u}\left(T_{N-1},L(T_{N-1})\right)-g(L(T_{N-1}))|^{2}\right]\\
=\frac{1}{S}\sum_{i=0}^{S-1}\left(\hat{u_{i}}\left(T_{N-1},L(T_{N-1})\right)-g_{i}(L(T_{N-1}))\right)^{2}
\end{gathered}
\end{equation}
We can now use a stochastic gradient descent (SGD) algorithm to optimize
the parameters, just as in the training of deep neural networks. After
the optimization, the initial option price and Delta can be obtained.
In our numerical examples, we use the Adam optimizer. For details
of the methodology, we refer to papers \cite{Weinan17,Jiequn17}.

As aforementioned, this forward solver is only suitable for pricing
European style options. The main advantages of this solver are as
follows:
\begin{itemize}
\item The Deltas are the direct outputs of this solver; while in Monte Carlo
simulation, Deltas are normally calculated by using shock and revaluation
method. Therefore, this method can save a large amount of computation
times, which is critical for trading desks especially during periods
when financial markets are very volatile. 
\item This solver can handle high dimensional problems, which can not be
handled by traditional PDE numerical method like finite difference
method or finite element method. Therefore, this new approach makes
it possible that some complex high-dimensional financial models like
Libor market model can be solved in acceptable time intervals.
\end{itemize}

\subsubsection{Stochastic Optimization Algorithms for Backward DNN Solver\label{subsec:BackwardSolver}}

In this subsection we present the new backward solver algorithm to
solve European option pricing which mainly follows the steps below.
Later, we discuss how the second step can be easily adjusted to meet
the needs of Bermudan option pricing.
\begin{itemize}
\item First, use Eq.(\ref{eq:Libor-SDE-Discrete}) to project Libor rates
forward from $\left(t_{0},L(t_{0})\right)$ to $\left(T_{N-1},L(T_{N-1})\right)$.
\end{itemize}
Second, use $\left(T_{N-1},L(T_{N-1})\right)$ obtained in previous
step to calculate the final discounted option payoff $\hat{u}\left(T_{N-1},L(T_{N-1})\right)=g(L(T_{N-1}))$, 

and then we use Eq.(\ref{eq:Option-Price-Approx-Back-Solver}) to
project discounted option price backward from $\left(T_{N-1},u\left(T_{N-1},L(T_{N-1})\right)\right)$
to $\left(t_{0},u(t_{0},L(t_{0}))\right)$ via 
\begin{equation}
\hat{u}\left(t_{i},L(t_{i})\right)=\hat{u}(t_{i+1},L(t_{i+1}))-\left(\nabla\hat{u}\sigma\right)\left(t_{i},L(t_{i})|\theta_{i}\right)\left(W\left(t_{i+1}\right)-W\left(t_{i}\right)\right)\label{eq:Back-flow}
\end{equation}
for $i=N-1,\ldots,0$.
\begin{itemize}
\item Third, in theory, $\hat{u}\left(t_{0},L(t_{0})\right)$ across all
the backward simulated samples should be identical. Therefore our
main idea is that after obtaining the neural network approximated
sample initial option prices $\hat{u}\left(t_{0},L(t_{0})\right)$,
we should minimize the expected loss function:
\begin{equation}
l\left(\theta\right)=E\left[|\hat{u}\left(t_{0},L(t_{0})\right)-E\left(\hat{u}\left(t_{0},L(t_{0})\right)\right)|^{2}\right]
\end{equation}
which is also the variance of $\hat{u}\left(t_{0},L(t_{0})\right)$.
When $S$ samples of Monte Carlo are generated, $l(\theta)$ can also
be re-written as: 
\begin{equation}
\begin{gathered}l\left(\theta\right)=E\left[|\hat{u}\left(t_{0},L(t_{0})\right)-E\left(\hat{u}\left(t_{0},L(t_{0})\right)\right)|^{2}\right]\\
=\frac{1}{S}\sum_{i=0}^{S-1}\left(\hat{u_{i}}\left(t_{0},L(t_{0})\right)-\frac{1}{S}\sum_{i=0}^{S-1}\hat{u_{i}}\left(t_{0},L(t_{0})\right)\right)^{2}
\end{gathered}
\end{equation}
After minimizing the loss function, $\frac{1}{S}\sum_{i=0}^{S-1}\hat{u_{i}}\left(t_{0},L(t_{0})\right)$
is defined as the initial discounted option price; and neural network
approximated $E\left(\nabla\hat{u}\left(t_{0},L(t_{0})\right)\right)$
can be used as the initial Libor Delta. The complete set of parameters
$\theta=\left\{ \theta_{1},\ldots,\theta_{N-1}\right\} $ are adjusted
in the process of approximating the option price gradients. The essential
difference lies in the fact that for backward DNN solver the initial
option price and gradients are not included in this parameter set.
Other than the backward projection, our backward solver shares a similar
structure in discretization and parameterization compared to the forward
solver which we discussed in prior two subsections. We aslo refer
to section 4.1 of paper \cite{Weinan17} for more detail. We use a
stochastic gradient descent (SGD) algorithm to optimize the parameters,
same approach adopted for the training of forward feed deep neural
networks. In our numerical examples, we continue to use the Adam optimizer. 
\end{itemize}
Figure 3.1 illustrates the architecture of the deep BSDE backward
solver with $m=N-1$.

\begin{figure}[h]
\includegraphics[width=1.2\textwidth]{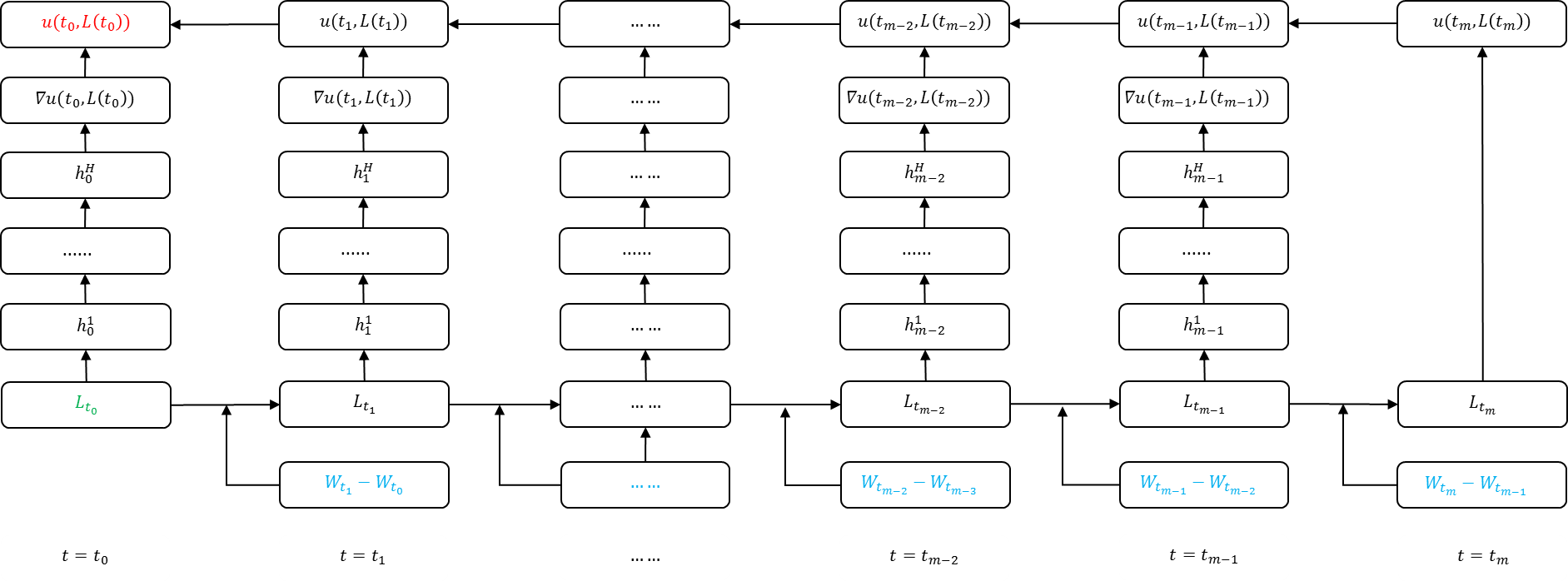}

\caption{Flowchart of backward BSDE solver}

\end{figure}

There are three types of connections in this network:
\begin{enumerate}
\item $(L(t_{i}),W(t_{i+1})-W(t_{i}))\rightarrow L(t_{i+1})$ is characterized
by (\ref{eq:Libor-SDE-Discrete}) using Euler/predictor-corrector
scheme. There are no parameters to be optimized in this type of connection.
\item $L(t_{i})\rightarrow h_{i}^{1}\rightarrow h_{i}^{2}\rightarrow\cdots\rightarrow h_{i}^{H}\rightarrow\nabla u(t_{i},L(t_{i}))$
is the multilayer feed forward neural network approximating the spatial
gradients at time $t=t_{i}.$ Here we have $H$ hidden layers $h_{i}^{1},\ldots,h_{i}^{H}$
for each time point. The weights $\theta_{i}$ of this sub-network
are the parameters we aim to optimize. Basically, $\theta_{i}$ decides
the linear/nonlinear transformation from input layer to hidden layer,
between hidden layers and from last hidden layer to gradient output,
and includes all the potential batch-normalization parameters involved
in the process as well. We will illustrate more in our simulation.
For more detail, we refer to Section 4.1 of paper \cite{Weinan17}.
\item $(u(t_{i+1},L(t_{i+1})),\nabla u(t_{i},L(t_{i})),W(t_{i+1})-W(t_{i}))\rightarrow u(t_{i},L(t_{i}))$
is the backward iteration that finally outputs an approximation of
$u(t_{0},L(t_{0}))$, completely characterized by (\ref{eq:Back-flow})
. There are no parameters to be optimized in this type of connections.
\end{enumerate}
The backward solver enjoys many same advantages of the forward solver,
which we discussed in the previous sections. Additionally, this backward
solver can not only to price European options, but also to price Bermudan
swaptions which we will soon discuss in Section 4. The option price
BSDE (\ref{eq:Option-BSDE}) is projected backward. When passing one
exercise date, the exercise information on this exercise date can
be easily used to update the value by the following formula:
\begin{equation}
u\left(t_{-},L(t_{-})\right)=max\left(u\left(t_{+},L(t_{+})\right),\;DiscIntrinsicValue(t)\right)
\end{equation}
where $t_{-}$ represents the time immediately before the exercise
time, $t_{+}$ represents the time immediately after the exercise
time. $DiscIntrinsicValue(t)$ represents discounted intrinsic value
for option at the exercise time, which depends on the type of option.
For example, the discounted intrinsic value for a receiver Bermudan
swaption at exercise time t is:

\begin{equation}
DiscIntrinsicValue(t)=\left(\sum_{i=q(t)}^{N-1}\left(K-L_{i}(t)\right)\tau_{i}\right)/\left(\prod_{j=0}^{q(t)-1}\left(1+L_{j}(t)\tau_{j}\right)\right)
\end{equation}
where $K$ is the fixed rate of the underlying swap, $q(t)$ is the
index of the first Libor rate which is not reset at exercise time
$t$, $N$ is the total number of Libor rates, $\tau$ is the accrual
time. Here we assume accrual time for fixed leg and float leg is the
same and exercise time t is part of the reset dates of Libor rates.

In more detail, suppose a Bermudan swaption has exercise dates on
$t_{k_{1},}\ldots,t_{k_{p}}$ (For a Bermudan swaption, all the exercise
dates are part of Libor reset dates) where the last exercise date
$t_{k_{p}}=T_{N-1}$. The only adjustment needed on exercise date
for the Bermudan swaption is to replace the Eq.(\ref{eq:Back-flow})
with the following backward projected evaluation:
\begin{equation}
\hat{u}\left(t_{i},L(t_{i})\right)=\begin{cases}
\hat{u}(t_{i+1},L(t_{i+1}))-\left(\nabla\hat{u}\sigma\right)\left(t_{i},L(t_{i})|\theta_{i}\right)\left(W\left(t_{i+1}\right)-W\left(t_{i}\right)\right)\\
\quad\quad\quad\quad\quad\quad\quad\quad\quad\quad\quad\quad\quad\quad\quad\quad i\notin\{k_{1,},\ldots,k_{p-1},k_{p}\}\\
\begin{gathered}max\{\hat{u}(t_{i+1},L(t_{i+1}))-\left(\nabla\hat{u}\sigma\right)\left(t_{i},L(t_{i})|\theta_{i}\right)\left(W\left(t_{i+1}\right)-W\left(t_{i}\right)\right),\\
DiscIntrinsicValue(t)\}
\end{gathered}
\\
\quad\quad\quad\quad\quad\quad\quad\quad\quad\quad\quad\quad\quad\quad\quad\quad i\in\{k_{1,},\ldots,k_{p-1}\}
\end{cases}\label{eq:Back-flow-Berm}
\end{equation}
Here, the exercise dates partition the time horizon into multiple
time intervals $(t_{0,}t_{k_{1}}),\ (t_{k_{1}},t_{k_{2}}),\ldots(t_{k_{p-1}},t_{k_{p}})$
and iteration formula within each time interval remains the same as
the European swaption case, but there is only one difference in the
exercise dates $t_{k_{1},}\ldots,t_{k_{p}}$ where the backward update
is now the max of (1) first order approximation of price function
with information carried backwards and (2) the current exercise value.

Between any two consecutive exercise dates, e.g. within $(t_{k_{i-1}},t_{k_{i}})$,
$\hat{u}$ is the approximation of holding value for that specific
sub-interval, i.e. $\hat{H_{i}}$ as in subsection\ref{subsec:From-European-to-Bermudan}.
The validity of this approximation is guaranteed by Eq. \ref{eq:hold-value-iteration}
which is a backward iteration to price the holding values. Similar
to the backward simulation approach in papers\cite{Longstaff01,Andersen98,Broadie97},
the backward BSDE solver actually still forward simulates the dynamics
of Libor rates. However, the application of Bellman dynamic programming
based on the simulated underlying Libor rate dynamics a reliable approximation
of holding value in a later period$(t_{k_{i}},t_{k_{i+1}})$ helps
to determine the holding value for the period$(t_{k_{i-1}},t_{k_{i}})$
right before, and hence the approximation of holding values for all
the simulated paths in a backward way is made feasible. Unlike the
conventional Monte Carlo pricer that may relies on exponentially growing
number of paths to evaluate the expected future value, our neural
network BSDE solver needs to use very limited number of sample paths
since it uses a parametric form gradient approximation which achieves
the efficient use of information across all sample scenarios. Compared
with classical approach in papers \cite{Longstaff01,Andersen98,Broadie97}
the neural network approach adopts a trainable and more complex form
of function approximation that leads to accuracy much efficiently. 

The capability of backward solver to price Bermudan swaption shoes
the potential application of DNN based BSDE solver in a wide range
of different Bermudan- style rate options, such as Bermudan swaptions,
Callable CMS spread options, Callable Bonds and Callable Range Accruals,
to name a few. 

In the next section, we will test the robustness of forward and backward
solvers under varying trade settings to price interest rate options
with respect to solver convergence, stability and accuracy.

\section{Numerical Results\label{sec:Numerical-Results}}

We show numerical examples using deep neural network BSDE solver to
price various interest rate options in this section. The first example
is to price co-terminal European swaptions using the forward solver.
The second example assumes a flat yield curve to price European swaptions
using forward and backward solvers respectively. The pricing differences
are very small between the two solvers. The third example uses a real
market yield curve instead of a flat yield curve. The pricing results
again show the differences between the two BSDE solvers are non-material.
The fourth example is to price a short maturity Bermudan swaption
pricing using the backward solver. The last example is for a long
maturity Bermudan swaption pricing by the backward solver. Each of
the numerical examples is performed with Python programming using
TensorFlow\footnote{An open-source library for data flow programming across a range of
tasks.}.

\subsection{Co-terminal European Swaption Pricing using Forward Solver \label{subsec:4.1 Co-terminal-European-Swaption-using-Forward-Solver}}

First, we use the forward solver to price European swaptions. Under
spot measure, we can build the BSDE solver by combining Eqs.(\ref{eq:Libor-SDE-Spot-Measure},
\ref{eq:swaption-PDE}, \ref{eq:BSDE} and \ref{eq:Option-BSDE}). 

The swaption, market inputs, and LMM set-up are listed as follows:
\begin{itemize}
\item The swaptions we use for testing are ATM co-terminal swaptions with
notional equal to 1 and terminal time $T_{10}=5.0722\;Yr$. The expiry
time is listed in the following table: 
\begin{table}[H]
\makebox[\textwidth][c]{%
\begin{tabular}{|c|c|c|c|c|c|c|c|c|c|c|}
\hline 
{\footnotesize{}T} & {\footnotesize{}$T_{0}$} & {\footnotesize{}$T_{1}$} & {\footnotesize{}$T_{2}$} & {\footnotesize{}$T_{3}$} & {\footnotesize{}$T_{4}$} & {\footnotesize{}$T_{5}$} & {\footnotesize{}$T_{6}$} & {\footnotesize{}$T_{7}$} & {\footnotesize{}$T_{8}$} & {\footnotesize{}$T_{9}$}\tabularnewline
\hline 
\hline 
{\footnotesize{}Expiry} & {\footnotesize{}0.0000} & {\footnotesize{}0.5028} & {\footnotesize{}1.0139} & {\footnotesize{}1.5167} & {\footnotesize{}2.0278} & {\footnotesize{}2.5333} & {\footnotesize{}3.0472} & {\footnotesize{}3.5528} & {\footnotesize{}4.0583} & {\footnotesize{}4.5639}\tabularnewline
\hline 
\end{tabular}}\caption{The expiries of co-terminal swaptions}
\end{table}
\item In order to price these swaptions, there are nine dynamic processes
in LMM corresponding to Libor rates $L_{i}(t),\;i=1,\cdots,9$. There
are nine Brownian motions driving these processes, and the model is
of nine factors. Since the first Libor rate $L_{0}(t)$ is already
reset at time $T_{0}=0$, there is no stochastic process driving this
rate.
\item The volatility term structure for each Libor rate $L_{i}(t)$ is a
hump-shaped function: $\xi_{i}(t,\,L(t))=\|\sigma_{i}(t,\,L(t))\|=L_{i}(t)\left(\left(a(T_{i}-t)+d\right)exp\left(-b(T_{i}-t)\right)+c\right)$.
The values of volatility parameters are listed in the following table:
\begin{table}[H]
\makebox[\textwidth][c]{%
\begin{tabular}{|c|c|c|c|c|}
\hline 
{\footnotesize{}Parameter} & {\footnotesize{}a} & {\footnotesize{}b} & {\footnotesize{}c} & {\footnotesize{}d}\tabularnewline
\hline 
\hline 
{\footnotesize{}Value} & {\footnotesize{}0.291} & {\footnotesize{}1.483} & {\footnotesize{}0.116} & {\footnotesize{}0.00001}\tabularnewline
\hline 
\end{tabular}}\caption{The values of vol parameters}
\end{table}
\item The correlation structure of Libor rates $L_{i}(t),\;i=1,\cdots,9$,
are defined by function: $\rho_{ij}=exp(-\beta|i-j|)$, where $\beta$
is set to be 0.5.
\item Yield curve is a flat zero curve with continuous compounding rate
4.00\%, from which the initial Libor rates are calculated, which are
listed in the following table: 
\begin{table}[H]
\makebox[\textwidth][c]{%
\begin{tabular}{|c|c|c|c|c|c|c|c|c|c|c|}
\hline 
{\footnotesize{}Libor Rate} & {\footnotesize{}$L_{0}$} & {\footnotesize{}$L_{1}$} & {\footnotesize{}$L_{2}$} & {\footnotesize{}$L_{3}$} & {\footnotesize{}$L_{4}$} & {\footnotesize{}$L_{5}$} & {\footnotesize{}$L_{6}$} & {\footnotesize{}$L_{7}$} & {\footnotesize{}$L_{8}$} & {\footnotesize{}$L_{9}$}\tabularnewline
\hline 
\hline 
{\footnotesize{}Value} & {\footnotesize{}4.0405\%} & {\footnotesize{}4.0412\%} & {\footnotesize{}4.0405\%} & {\footnotesize{}4.0412\%} & {\footnotesize{}4.0407\%} & {\footnotesize{}4.0414\%} & {\footnotesize{}4.0407\%} & {\footnotesize{}4.0407\%} & {\footnotesize{}4.0407\%} & {\footnotesize{}4.0409\%}\tabularnewline
\hline 
\end{tabular}}\caption{The initial Libor rates.}
\end{table}
\end{itemize}
We use predictor-corrector method to discretize Eq.(\ref{eq:Libor-SDE-Spot-Measure})
and the time step is one month. We show some of the projected shapes
of the yield curves Figure 4.1. From the figure, we can observe that
this 9-factor LMM can generate rich shapes of yield curves: inverted,
flat, and upward. 

\begin{figure}[H]
\makebox[\textwidth][c]{\includegraphics[width=0.8\textwidth]{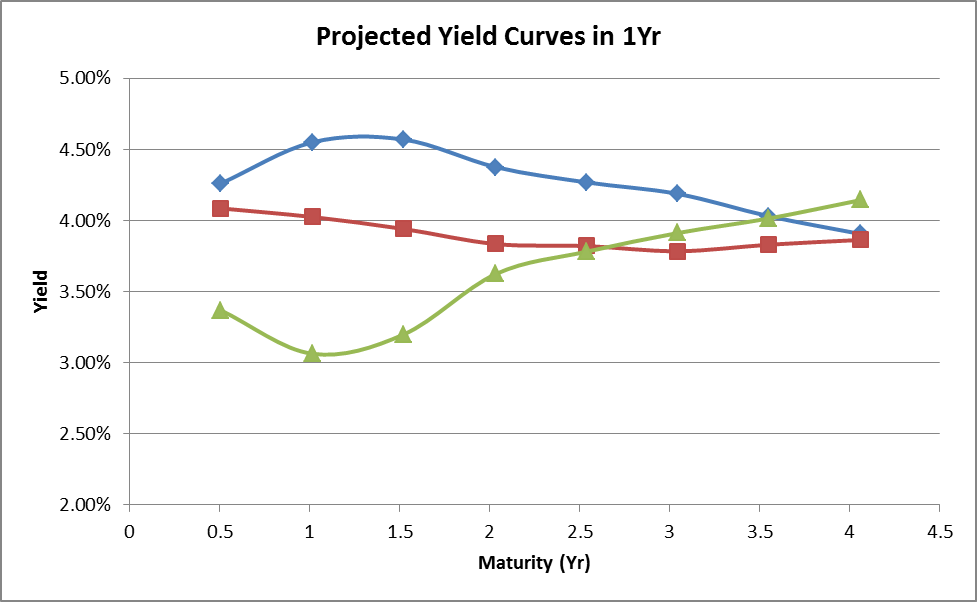}}

\caption{Projected yield curves in 1Yr}
\end{figure}

The deep neural network setting is: each of the sub neural network
approximating gradient $\nabla u$ consists of 4 layers (1 input layer
{[}d-dimensional{]}, 2 hidden layers {[}both d+10-dimensional{]},
and 1 output layer {[}d-dimensional{]}). In this case d = 10. The
inputs of this sub-neural network are Libor rates $L_{i}(t),\;i=0,\cdots,9$
and its outputs are the gradient of option prices with respect to
the inputs. In the test, we run 10,000 optimization iterations and
use 4096 Monte Carlo samples for optimization. 

We compare the results with the the results from Quantlib for sort
of benchmarking. The method used in QuantLib is conventional Monte
Carlo simulation with 50,000 scenarios. In the test, we take notional
of swaption to be 1. The results are very close and their relative
differences are well within 0.5\%. The details are listed in the Table
\ref{tab:NPVs-Coterminal-Swaption-QuantLib} through Table \ref{tab:Diff-Coterminal-Swaption}. 

\begin{table}[H]
\makebox[\textwidth][c]{%
\begin{tabular}{|c|c|c|c|c|c|c|}
\hline 
{\small{}Expiry\textbackslash{}Tenor} & {\small{}1.0139} & {\small{}2.0250} & {\small{}3.0444} & {\small{}3.5556} & {\small{}4.0583} & {\small{}4.5694}\tabularnewline
\hline 
\hline 
{\small{}0.5028} &  &  &  &  &  & {\small{}0.004050}\tabularnewline
\hline 
{\small{}1.0139} &  &  &  &  & {\small{}0.005312} & \tabularnewline
\hline 
{\small{}1.5167} &  &  &  & {\small{}0.005879} &  & \tabularnewline
\hline 
{\small{}2.0278} &  &  & {\small{}0.006034} &  &  & \tabularnewline
\hline 
{\small{}3.0472} &  & {\small{}0.005394} &  &  &  & \tabularnewline
\hline 
{\small{}4.0583} & {\small{}0.003554} &  &  &  &  & \tabularnewline
\hline 
\end{tabular}}\caption{The NPVs of ATM co-terminal swaptions from QuantLib \label{tab:NPVs-Coterminal-Swaption-QuantLib}}
\end{table}

\begin{table}[H]
\makebox[\textwidth][c]{%
\begin{tabular}{|c|c|c|c|c|c|c|}
\hline 
{\small{}Expiry\textbackslash{}Tenor} & {\small{}1.0139} & {\small{}2.0250} & {\small{}3.0444} & {\small{}3.5556} & {\small{}4.0583} & {\small{}4.5694}\tabularnewline
\hline 
\hline 
{\small{}0.5028} &  &  &  &  &  & {\small{}0.004068}\tabularnewline
\hline 
{\small{}1.0139} &  &  &  &  & {\small{}0.005328} & \tabularnewline
\hline 
{\small{}1.5167} &  &  &  & {\small{}0.005883} &  & \tabularnewline
\hline 
{\small{}2.0278} &  &  & {\small{}0.006024} &  &  & \tabularnewline
\hline 
{\small{}3.0472} &  & {\small{}0.005385} &  &  &  & \tabularnewline
\hline 
{\small{}4.0583} & {\small{}0.003566} &  &  &  &  & \tabularnewline
\hline 
\end{tabular}}\caption{The NPVs of ATM co-terminal swaptions from forward solver \label{tab:NPVs-Coterminal-Swaption-NN}}
\end{table}

\begin{table}[H]
\makebox[\textwidth][c]{%
\begin{tabular}{|c|c|c|c|c|c|c|}
\hline 
{\small{}Expiry\textbackslash{}Tenor} & {\small{}1.0139} & {\small{}2.0250} & {\small{}3.0444} & {\small{}3.5556} & {\small{}4.0583} & {\small{}4.5694}\tabularnewline
\hline 
\hline 
{\small{}0.5028} &  &  &  &  &  & {\small{}0.44\%}\tabularnewline
\hline 
{\small{}1.0139} &  &  &  &  & {\small{}0.31\%} & \tabularnewline
\hline 
{\small{}1.5167} &  &  &  & {\small{}0.06\%} &  & \tabularnewline
\hline 
{\small{}2.0278} &  &  & {\small{}-0.16\%} &  &  & \tabularnewline
\hline 
{\small{}3.0472} &  & {\small{}-0.17\%} &  &  &  & \tabularnewline
\hline 
{\small{}4.0583} & {\small{}0.35\%} &  &  &  &  & \tabularnewline
\hline 
\end{tabular}}\caption{The difference of NPVs of ATM co-terminal swaptions between QuantLib
and forward solver \label{tab:Diff-Coterminal-Swaption}}
\end{table}

Another benefit of the DNN-based forward solver is that the first
order sensitivity Delta can be direct outputs from the valuation process,
which saves a large amount of computational time. While in conventional
Monte Carlo simulation, the Delta is calculated by shock and revalue
method, Table \ref{tab:Diff-Delta-Swaption-Forward} shows their comparison
for swaption (expiry: 1.5167Yr, tenor: 3.5556Yr). From the table,
we conclude the results are very close: for Libor rates $L_{0},\;L_{1},\;and\quad L_{2}$,
they are mainly used to discount cash flows and have very small exposure
to the swaption price. Hence, Deltas w.r.t those rates are very small
and their relative differences are relatively large.

\begin{table}[H]
\makebox[\textwidth][c]{%
\begin{tabular}{|c|c|c|c|c|c|c|c|c|c|c|}
\hline 
{\small{}Libor Rate} & {\small{}$L_{0}$} & {\small{}$L_{1}$} & {\small{}$L_{2}$} & {\small{}$L_{3}$} & {\small{}$L_{4}$} & {\small{}$L_{5}$} & {\small{}$L_{6}$} & {\small{}$L_{7}$} & {\small{}$L_{8}$} & {\small{}$L_{9}$}\tabularnewline
\hline 
\hline 
{\small{}QuantLib} & {\small{}-0.0029} & {\small{}-0.0029} & {\small{}-0.0029} & {\small{}-0.2234} & {\small{}-0.2131} & {\small{}-0.2144} & {\small{}-0.2096} & {\small{}-0.2084} & {\small{}-0.2075} & {\small{}-0.2074}\tabularnewline
\hline 
{\small{}Forward solver} & {\small{}0.0007} & {\small{}-0.0028} & {\small{}-0.0018} & {\small{}-0.2249} & {\small{}-0.2127} & {\small{}-0.2149} & {\small{}-0.2099} & {\small{}-0.2095} & {\small{}-0.2083} & {\small{}-0.2077}\tabularnewline
\hline 
{\small{}Relative\_Diff} & {\small{}-125.14\%} & {\small{}-2.97\%} & {\small{}-37.31\%} & {\small{}0.67\%} & {\small{}-0.19\%} & {\small{}0.25\%} & {\small{}0.12\%} & {\small{}0.51\%} & {\small{}0.39\%} & {\small{}0.15\%}\tabularnewline
\hline 
\end{tabular}}\caption{The comparison of Delta between QuantLib and DNN \label{tab:Diff-Delta-Swaption-Forward}}
\end{table}

Overall, we found that deep neural network based BSDE solver is effective
for derivative pricing under LMM. One of the great benefits is that
it can generate the first order sensitivity directly.

\subsection{Comparison of Forward Solver and Backward Solver in European Swaption
Pricing}

In this subsection, we use both DNN-based forward and backward solvers
to price European swaptions and then we compare their performance
from the perspective of accuracy and convergence. The Libor market
model set-up, market information, BSDEs, and Neural Network setting
are the same as in Section \ref{subsec:4.1 Co-terminal-European-Swaption-using-Forward-Solver}.
The European swaption we use for testing is ATM with expiry $T_{2}=1.0139Yr$
and terminal time for underlying swap $T_{10}=5.0722Yr$, and its
notional is 1.

Table \ref{tab:Swaption-Forward-Solver} and Table \ref{tab:Swaption-Backward-Solver}
present the calculation of the mean of $u\left(t_{0}\right)$, the
standard deviation of $u\left(t_{0}\right)$, the mean of the loss
function $l(\theta)$, the standard deviation of the loss function
$l(\theta)$, and the run time in seconds needed to calculate one
realization of $u\left(t_{0},L(t_{0})\right)$ against number of optimization
iterations $m\in\left\{ 2000,4000,6000\right\} $ based on 5 independent
runs and 4096 Monte Carlo samples with monthly time step. From the
tables, we conclude the following:
\begin{itemize}
\item The approximated Europen swaption prices from the two solvers converge
very fast to around 0.00532. For forward solver, the price converges
with 4,000 iterations while it needs only 2,000 iterations for backward
solver.
\item The tests run on one V100 GPU in DGX-1 server, the speed is similar
for forward and backward solver.
\end{itemize}
\begin{table}[H]
\makebox[\textwidth][c]{%
\begin{tabular}{|>{\centering}p{2cm}||>{\centering}m{2cm}||>{\centering}p{2cm}||>{\centering}p{2cm}||>{\centering}p{2cm}||>{\centering}p{2cm}|}
\hline 
{\small{}Number of Iteration Steps m} & {\small{}Mean of }{\footnotesize{}$u(0)$} & {\small{}STD of }{\footnotesize{}$u(0)$} & {\small{}Mean of the Loss Function} & {\small{}STD of the Loss Function} & {\small{}Runtime in sec. for one run}\tabularnewline
\hline 
\hline 
{\small{}2000} & {\small{}0.005330} & {\small{}4.39E-06} & {\small{}1.70E-05} & {\small{}4.31E-06} & {\small{}37}\tabularnewline
\hline 
\hline 
{\small{}4000} & {\small{}0.005330} & {\small{}1.14E-06} & {\small{}1.67E-06} & {\small{}4.48E-07} & {\small{}74}\tabularnewline
\hline 
\hline 
{\small{}6000} & {\small{}0.005329} & {\small{}1.53E-06} & {\small{}1.18E-06} & {\small{}2.90E-08} & {\small{}106}\tabularnewline
\hline 
\end{tabular}}\caption{Numerical simulaions for the deep BSDE forward solver for European
swaption \label{tab:Swaption-Forward-Solver}}
\end{table}

\begin{table}[H]
\makebox[\textwidth][c]{%
\begin{tabular}{|>{\centering}p{2cm}||>{\centering}m{2cm}||>{\centering}p{2cm}||>{\centering}p{2cm}||>{\centering}p{2cm}||>{\centering}p{2cm}|}
\hline 
{\small{}Number of Iteration Steps m} & {\small{}Mean of u} & {\small{}STD of u} & {\small{}Mean of the Loss Function} & {\small{}STD of the Loss Function} & {\small{}Runtime in sec. for one run}\tabularnewline
\hline 
\hline 
{\small{}2000} & {\small{}0.005318} & {\small{}3.31E-06} & {\small{}3.39E-06} & {\small{}1.13E-08} & {\small{}38}\tabularnewline
\hline 
\hline 
{\small{}4000} & {\small{}0.005315} & {\small{}7.94E-06} & {\small{}3.31E-06} & {\small{}5.55E-09} & {\small{}73}\tabularnewline
\hline 
\hline 
{\small{}6000} & {\small{}0.005322} & {\small{}7.17E-06} & {\small{}3.30E-06} & {\small{}1.01E-08} & {\small{}109}\tabularnewline
\hline 
\end{tabular}}\caption{Numerical simulaions for the deep BSDE Backward solver for European
swaption \label{tab:Swaption-Backward-Solver}}
\end{table}

\begin{figure}[H]
\makebox[\textwidth][c]{\includegraphics[width=0.8\textwidth]{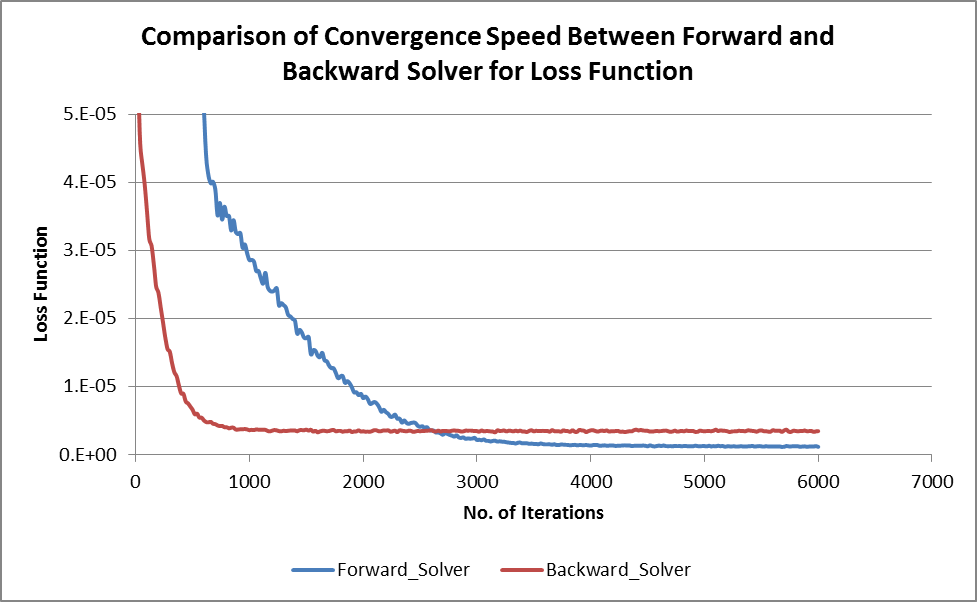}}

\caption{Comparison of convergence speed between forward and backward solver
for loss function\label{fig:Sec3.3 Comparison-of-convergence-Los-Function}}
\end{figure}

\begin{figure}[H]
\makebox[\textwidth][c]{\includegraphics[width=0.8\textwidth]{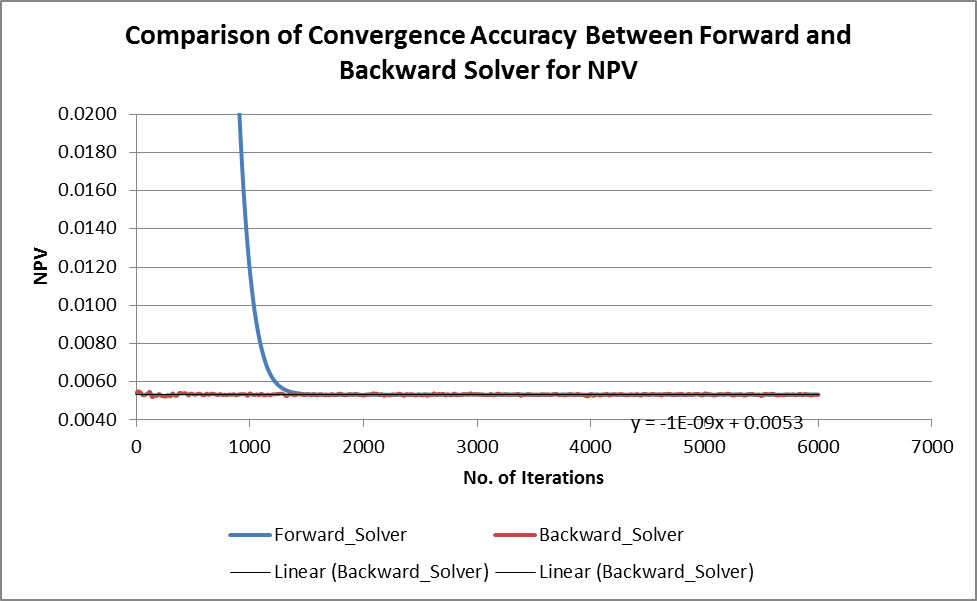}}

\caption{Comparison of convergence speed between forward and backward solver
for NPV \label{fig:Sec3.3 Comparison-of-convergence-NPV}}
\end{figure}

\begin{figure}[H]
\makebox[\textwidth][c]{\includegraphics[width=0.8\textwidth]{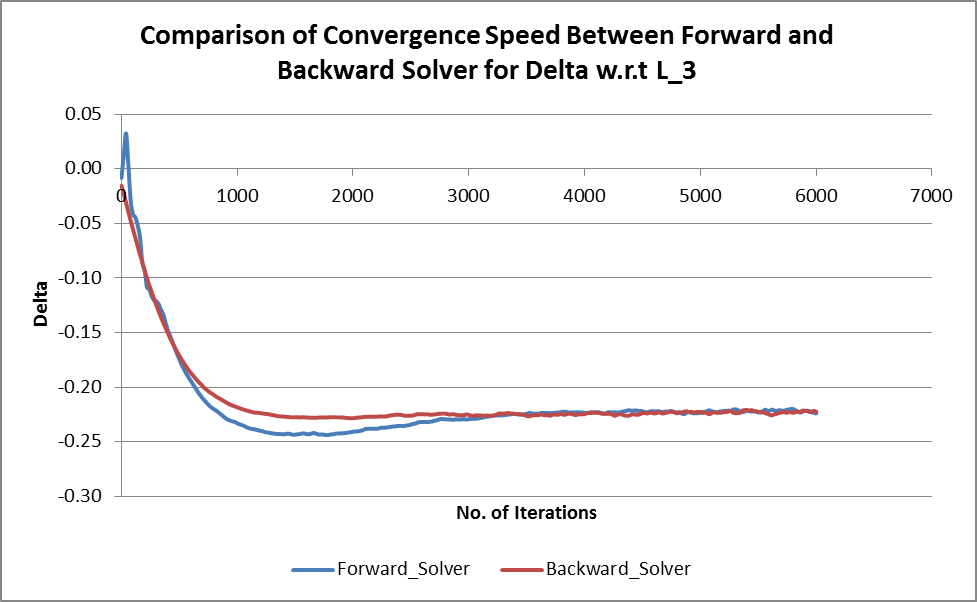}}

\caption{Comparison of convergence speed between forward and backward solver
for Delta w.r.t $L_{3}$ \label{fig:Sec3.3 Comparison-of-convergence-L-3}}
\end{figure}

\begin{figure}[H]
\makebox[\textwidth][c]{\includegraphics[width=0.8\textwidth]{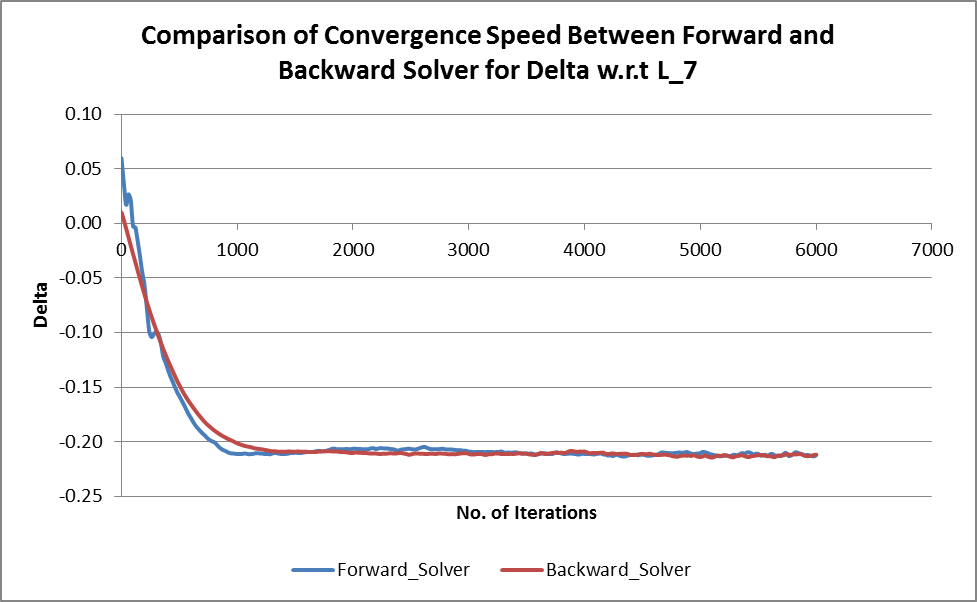}}

\caption{Comparison of convergence speed between forward and backward solver
for Delta w.r.t $L_{7}$ \label{fig:Sec3.3 Comparison-of-convergence-L-7}}
\end{figure}

\noindent Figure \ref{fig:Sec3.3 Comparison-of-convergence-Los-Function}
through Figure \ref{fig:Sec3.3 Comparison-of-convergence-L-7} show
the convergence speed and accuracy of NPVs and Deltas from the two
solvers. From the figures, we conclude that
\begin{itemize}
\item the accuracy of the two solvers are are quite close, but the convergence
speed of the back solver is faster than that of the forward solver;
\item the loss function of the backward solver is above that of the forward
solver because they represent two different targets. The loss function
of the backward solver represents the variance of $u\left(t_{0},L(t_{0})\right)$
while the loss function of the forward solver represents mean of square
of difference between the projected terminal payoffs and the actual
terminal payoffs. It is clear that the performance of a BSDE solver
closely depends the on the loss function imposed for convergence.
\end{itemize}

\subsection{Comparison of Forward Solver and Backward Solver in Co-terminal European
Swaption Pricing \label{subsec:4.3-Comparison-of-Forward-Backward-Co-Terminal}}

In this subsection, we use a real market yield curve on January 3,
2017, for interest rate option pricing. The curve is shown in the
following figure: 
\begin{figure}[H]
\makebox[\textwidth][c]{\includegraphics[width=0.8\textwidth]{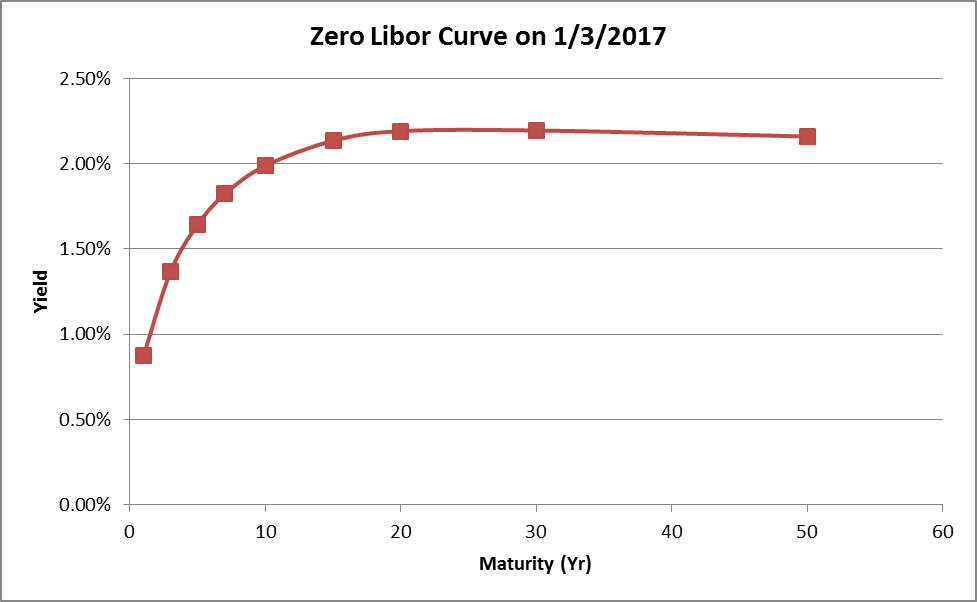}}

\caption{Zero Libor curve on business date 1/3/2017 \label{fig:Sec3.5 Real Libor curve}}
\end{figure}
.

\noindent It is an upward yield curve, used for the calculation of
the initial Libor rates, which are listed in the Table 11: 
\begin{table}[H]
\makebox[\textwidth][c]{%
\begin{tabular}{|c|c|c|c|c|c|c|c|c|c|c|}
\hline 
{\footnotesize{}Libor Rate} & {\footnotesize{}$L_{0}$} & {\footnotesize{}$L_{1}$} & {\footnotesize{}$L_{2}$} & {\footnotesize{}$L_{3}$} & {\footnotesize{}$L_{4}$} & {\footnotesize{}$L_{5}$} & {\footnotesize{}$L_{6}$} & {\footnotesize{}$L_{7}$} & {\footnotesize{}$L_{8}$} & {\footnotesize{}$L_{9}$}\tabularnewline
\hline 
\hline 
{\footnotesize{}Value} & {\small{}1.1842\%} & {\small{}1.1843\%} & {\small{}1.9085\%} & {\small{}1.9087\%} & {\small{}1.9067\%} & {\small{}1.9035\%} & {\small{}2.3785\%} & {\small{}2.3826\%} & {\small{}2.3826\%} & {\small{}2.3827\%}\tabularnewline
\hline 
\end{tabular}}\caption{The initial Libor rates on 1/3/2017.}
\end{table}
.

\noindent Except the yield curve, all the other information such as
swaption data, market data, LMM set-up, and neural network set-up
are the same as those contained in Section \ref{subsec:4.1 Co-terminal-European-Swaption-using-Forward-Solver}.
We mainly compare three numerical results: 
\begin{itemize}
\item Result from forward solver.
\item Result from backward solver.
\item Result from QuantLib.
\end{itemize}
\noindent The numerical results include NPVs of ATM co-terminal European
swaptions and direct outputs of Deltas for one swaption (expiry: 1.5167Yr,
tenor: 3.5556Yr). The results are listed in Table \ref{tab:NPVs-Coterminal-Swaption-QuantLib-Upward-Yield-Curve}
through Table \ref{tab:Diff-Delta-Swaption-NN-Up-Yield-Curve}. Checking
the results, we conclude the following:
\begin{itemize}
\item All the NPVs are very close and their relative differences to QuantLib
are all within 0.5\%. 
\item All the Detlas are very close except those w.r.t. Libor rates $L_{0},\;L_{1},\;and\quad L_{2}$,
which are mainly used to discount cash flows and have very small exposure
to the swaption price. Hence, Deltas w.r.t. these three Libor rates
are very small and close to 0 although their relative differences
are much larger. 
\item Using QuantLib as the base, the forward solver is slightly more accurate
than the backward solver. Relative differences of Delta from the forward
solver are mostly within 0.5\% range, while those of the backward
solver are mostly within 1.0\%.
\end{itemize}
Therefore, this comparison helps us to conclude that DNN-based solver
is accurate to price swaptions under distinct initial yield curves.

\begin{table}[H]
\makebox[\textwidth][c]{%
\begin{tabular}{|c|c|c|c|c|c|c|}
\hline 
{\small{}Expiry\textbackslash{}Tenor} & {\small{}1.0139} & {\small{}2.0250} & {\small{}3.0444} & {\small{}3.5556} & {\small{}4.0583} & {\small{}4.5694}\tabularnewline
\hline 
\hline 
{\small{}0.5028} &  &  &  &  &  & {\small{}0.002107}\tabularnewline
\hline 
{\small{}1.0139} &  &  &  &  & {\small{}0.002933} & \tabularnewline
\hline 
{\small{}1.5167} &  &  &  & {\small{}0.003339} &  & \tabularnewline
\hline 
{\small{}2.0278} &  &  & {\small{}0.003550} &  &  & \tabularnewline
\hline 
{\small{}3.0472} &  & {\small{}0.003482} &  &  &  & \tabularnewline
\hline 
{\small{}4.0583} & {\small{}0.002316} &  &  &  &  & \tabularnewline
\hline 
\end{tabular}}\caption{The NPVs of ATM co-terminal swaptions from QuantLib using initial
upward yield curve \label{tab:NPVs-Coterminal-Swaption-QuantLib-Upward-Yield-Curve}}
\end{table}

\begin{table}[H]
\makebox[\textwidth][c]{%
\begin{tabular}{|c|c|c|c|c|c|c|}
\hline 
{\small{}Expiry\textbackslash{}Tenor} & {\small{}1.0139} & {\small{}2.0250} & {\small{}3.0444} & {\small{}3.5556} & {\small{}4.0583} & {\small{}4.5694}\tabularnewline
\hline 
\hline 
{\small{}0.5028} &  &  &  &  &  & {\small{}0.002114}\tabularnewline
\hline 
{\small{}1.0139} &  &  &  &  & {\small{}0.002948} & \tabularnewline
\hline 
{\small{}1.5167} &  &  &  & {\small{}0.003337} &  & \tabularnewline
\hline 
{\small{}2.0278} &  &  & {\small{}0.003542} &  &  & \tabularnewline
\hline 
{\small{}3.0472} &  & {\small{}0.003479} &  &  &  & \tabularnewline
\hline 
{\small{}4.0583} & {\small{}0.002326} &  &  &  &  & \tabularnewline
\hline 
\end{tabular}}\caption{The NPVs of ATM co-terminal swaptions from forward solver using initial
upward yield curve \label{tab:NPVs-Coterminal-Swaption-Fwd-Solver-Upward-Yield-Curve}}
\end{table}

\begin{table}[H]
\makebox[\textwidth][c]{%
\begin{tabular}{|c|c|c|c|c|c|c|}
\hline 
{\small{}Expiry\textbackslash{}Tenor} & {\small{}1.0139} & {\small{}2.0250} & {\small{}3.0444} & {\small{}3.5556} & {\small{}4.0583} & {\small{}4.5694}\tabularnewline
\hline 
\hline 
{\small{}0.5028} &  &  &  &  &  & {\small{}0.35\%}\tabularnewline
\hline 
{\small{}1.0139} &  &  &  &  & {\small{}0.51\%} & \tabularnewline
\hline 
{\small{}1.5167} &  &  &  & {\small{}-0.07\%} &  & \tabularnewline
\hline 
{\small{}2.0278} &  &  & {\small{}-0.23\%} &  &  & \tabularnewline
\hline 
{\small{}3.0472} &  & {\small{}-0.07\%} &  &  &  & \tabularnewline
\hline 
{\small{}4.0583} & {\small{}0.42\%} &  &  &  &  & \tabularnewline
\hline 
\end{tabular}}\caption{The difference of NPVs of ATM co-terminal swaptions between QuantLib
and forward solver using initial upward yield curve \label{tab:Diff-Coterminal-Swaption-Fwd-Solver-Upward-Yield-Curve}}
\end{table}

\begin{table}[H]
\makebox[\textwidth][c]{%
\begin{tabular}{|c|c|c|c|c|c|c|}
\hline 
{\small{}Expiry\textbackslash{}Tenor} & {\small{}1.0139} & {\small{}2.0250} & {\small{}3.0444} & {\small{}3.5556} & {\small{}4.0583} & {\small{}4.5694}\tabularnewline
\hline 
\hline 
{\small{}0.5028} &  &  &  &  &  & {\small{}0.002104}\tabularnewline
\hline 
{\small{}1.0139} &  &  &  &  & {\small{}0.002940} & \tabularnewline
\hline 
{\small{}1.5167} &  &  &  & {\small{}0.003335} &  & \tabularnewline
\hline 
{\small{}2.0278} &  &  & {\small{}0.003539} &  &  & \tabularnewline
\hline 
{\small{}3.0472} &  & {\small{}0.003475} &  &  &  & \tabularnewline
\hline 
{\small{}4.0583} & {\small{}0.002324} &  &  &  &  & \tabularnewline
\hline 
\end{tabular}}\caption{The NPVs of ATM co-terminal swaptions from backward solver using initial
upward yield curve \label{tab:NPVs-Coterminal-Swaption-Back-Solver-Upward-Yield-Curve}}
\end{table}

\begin{table}[H]
\makebox[\textwidth][c]{%
\begin{tabular}{|c|c|c|c|c|c|c|}
\hline 
{\small{}Expiry\textbackslash{}Tenor} & {\small{}1.0139} & {\small{}2.0250} & {\small{}3.0444} & {\small{}3.5556} & {\small{}4.0583} & {\small{}4.5694}\tabularnewline
\hline 
\hline 
{\small{}0.5028} &  &  &  &  &  & {\small{}-0.15\%}\tabularnewline
\hline 
{\small{}1.0139} &  &  &  &  & {\small{}0.24\%} & \tabularnewline
\hline 
{\small{}1.5167} &  &  &  & {\small{}-0.13\%} &  & \tabularnewline
\hline 
{\small{}2.0278} &  &  & {\small{}-0.31\%} &  &  & \tabularnewline
\hline 
{\small{}3.0472} &  & {\small{}-0.20\%} &  &  &  & \tabularnewline
\hline 
{\small{}4.0583} & {\small{}0.35\%} &  &  &  &  & \tabularnewline
\hline 
\end{tabular}}\caption{The difference of NPVs of ATM co-terminal swaptions between QuantLib
and backward solver using initial upward yield curve \label{tab:Diff-Coterminal-Swaption-Back-Solver-Upward-Yield-Curve}}
\end{table}

\begin{table}[H]
\makebox[\textwidth][c]{%
\begin{tabular}{|c|c|c|c|c|c|c|c|c|c|c|}
\hline 
{\small{}Libor Rate} & {\small{}$L_{0}$} & {\small{}$L_{1}$} & {\small{}$L_{2}$} & {\small{}$L_{3}$} & {\small{}$L_{4}$} & {\small{}$L_{5}$} & {\small{}$L_{6}$} & {\small{}$L_{7}$} & {\small{}$L_{8}$} & {\small{}$L_{9}$}\tabularnewline
\hline 
\hline 
{\small{}QuantLib} & {\small{}-0.0017} & {\small{}-0.0017} & {\small{}-0.0016} & {\small{}-0.2356} & {\small{}-0.2266} & {\small{}-0.2291} & {\small{}-0.2240} & {\small{}-0.2242} & {\small{}-0.2250} & {\small{}-0.2324}\tabularnewline
\hline 
{\small{}Forward solver} & {\small{}0.0010} & {\small{}-0.0031} & {\small{}-0.0012} & {\small{}-0.2363} & {\small{}-0.2262} & {\small{}-0.2303} & {\small{}-0.2236} & {\small{}-0.2250} & {\small{}-0.2254} & {\small{}-0.2284}\tabularnewline
\hline 
{\small{}Relative\_Diff\_1} & {\small{}-160.76\%} & {\small{}79.71\%} & {\small{}-24.44\%} & {\small{}0.29\%} & {\small{}-0.17\%} & {\small{}0.54\%} & {\small{}-0.19\%} & {\small{}0.34\%} & {\small{}0.17\%} & {\small{}-1.73\%}\tabularnewline
\hline 
{\small{}Backward solver} & {\small{}0.0005} & {\small{}-0.0015} & {\small{}-0.0018} & {\small{}-0.2371} & {\small{}-0.2286} & {\small{}-0.2302} & {\small{}-0.2261} & {\small{}-0.2250} & {\small{}-0.2261} & {\small{}-0.2281}\tabularnewline
\hline 
Relative\_Diff\_2 & {\small{}-130.82\%} & {\small{}-13.12\%} & {\small{}9.94\%} & {\small{}0.63\%} & {\small{}0.90\%} & {\small{}0.47\%} & {\small{}0.94\%} & {\small{}0.35\%} & {\small{}0.49\%} & {\small{}-1.85\%}\tabularnewline
\hline 
\end{tabular}}\caption{The comparison of Deltas between QuantLib and deep neural network
based solvers. \label{tab:Diff-Delta-Swaption-NN-Up-Yield-Curve}}
\end{table}

\subsection{Short Maturity Bermudan Swaption Pricing using Backward Solver \label{4.4-subsec:Short-Term-Bermudan}}

\begin{figure}[H]
\makebox[\textwidth][c]{\includegraphics[width=0.8\textwidth]{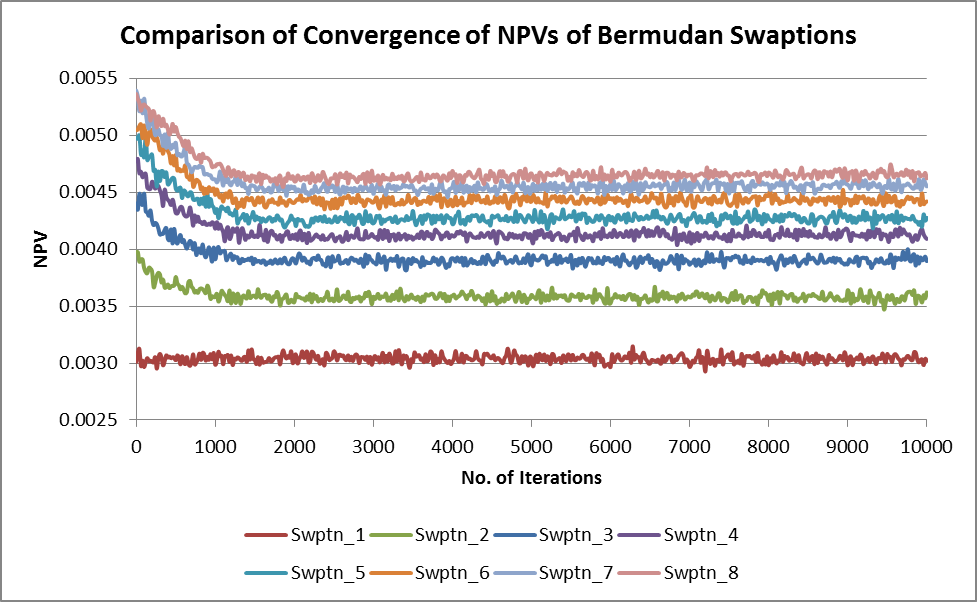}}

\caption{Convergence test of backward solver for NPVs of short term Bermudan
swaptions \label{fig:Sec4.4 Comparison-of-convergence-NPV-Bermudan-Swptn}}
\end{figure}

\begin{figure}[H]
\makebox[\textwidth][c]{\includegraphics[width=0.8\textwidth]{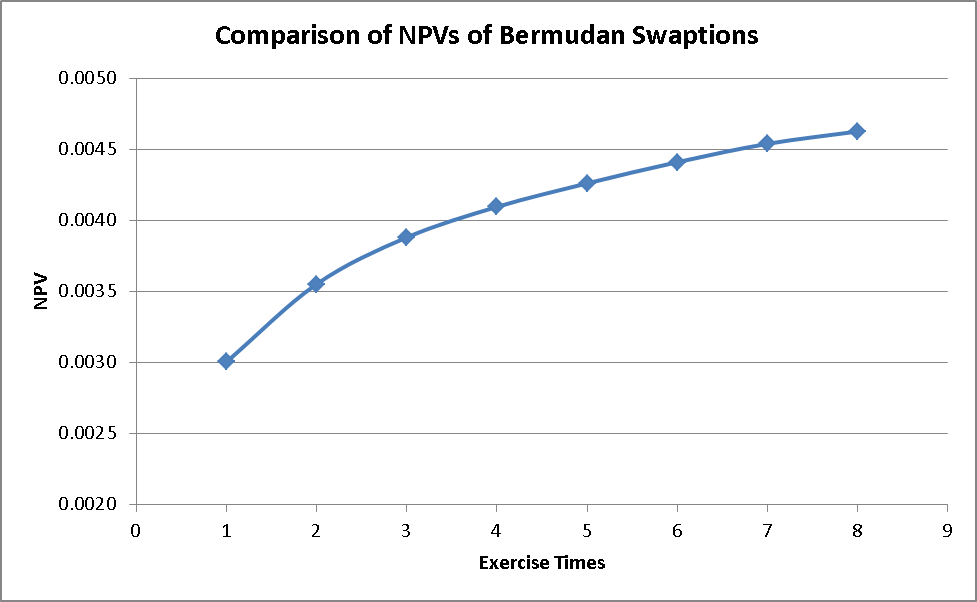}}

\caption{Comparison of NPVs of short term Bermudan swaptions with different
exercise times \label{fig:Sec4.4 Comparison-of-NPV-Diff-Bermudan-Swptn}}
\end{figure}

\begin{figure}[H]
\makebox[\textwidth][c]{\includegraphics[width=0.8\textwidth]{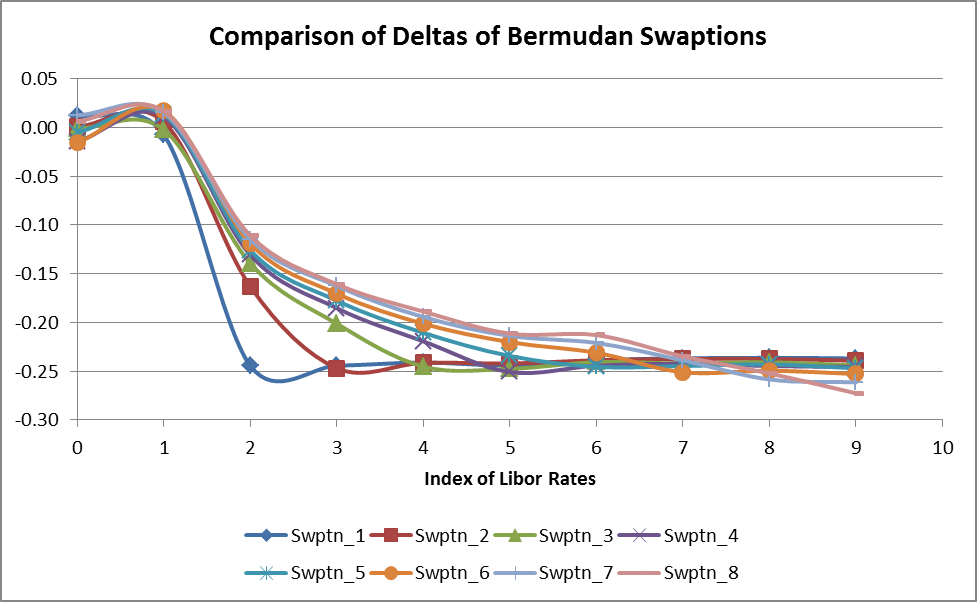}}

\caption{Comparison of Deltas of short term Bermudan swaptions \label{fig:Sec4.4 Comparison-of-Deltas-Bermudan-Swptn}}
\end{figure}

In this subsection, we use backward solver to price Bermudan swaptions
with short maturity. The Libor market model set-up, market information,
BSDEs and Neural Network setting are the same as those in Section
\ref{subsec:4.3-Comparison-of-Forward-Backward-Co-Terminal}. A total
of 4096 simulated samples/paths are used for optimization. The Bermudan
swaptions we use for testing are listed in Table 18.
\begin{itemize}
\item The exercise times of Bermudan swaptions are listed in the following
table 
\begin{table}[H]
\makebox[\textwidth][c]{%
\begin{tabular}{|c|c|c|c|c|c|c|c|c|}
\hline 
{\footnotesize{}T} & {\footnotesize{}$T_{2}$} & {\footnotesize{}$T_{3}$} & {\footnotesize{}$T_{4}$} & {\footnotesize{}$T_{5}$} & {\footnotesize{}$T_{6}$} & {\footnotesize{}$T_{7}$} & {\footnotesize{}$T_{8}$} & {\footnotesize{}$T_{9}$}\tabularnewline
\hline 
\hline 
{\footnotesize{}Expiry} & {\footnotesize{}1.0139} & {\footnotesize{}1.5167} & {\footnotesize{}2.0278} & {\footnotesize{}2.5333} & {\footnotesize{}3.0472} & {\footnotesize{}3.5528} & {\footnotesize{}4.0583} & {\footnotesize{}4.5639}\tabularnewline
\hline 
\end{tabular}}\caption{The exercise time of short term Bermudan swaption}
\end{table}
.
\item There are 8 Bermudan swaptions in total: Bermudan swaption 1 (swaption
1) contains one exercise time which is $T_{2}$; Bermudan swaption
2 (swaption 2) contains two exercise time which is $T_{2}$, $T_{3}$;
... ; Bermudan swaption 8 (swaption 8) contains eight exercise time
which is $T_{2}$, $T_{3}$, ..., $T_{9}$.
\item Their underlying swaps have the same terminal time which is 5.0722
Yr. 
\item They have the same notional, which is 1 and they have the same strike,
which is 2.1442\%.
\end{itemize}
Table \ref{tab:Sec4.4: The-NPVs-of-Bermudan-Swptn-1}, Table \ref{tab:Sec4.4: The-Deltas-of-Bermudan-Swptn},
and Figure \ref{fig:Sec4.4 Comparison-of-convergence-NPV-Bermudan-Swptn}
through Figure \ref{fig:Sec4.4 Comparison-of-Deltas-Bermudan-Swptn}
show the NPVs and Deltas of Bermudan swaptions. From the results,
we have the following observations:
\begin{itemize}
\item The convergence of NPVs is steady. With increasing number of exercise
times, the convergence speed slows down.
\item With the increase of exercise times, NPVs also increase quite smoothly
with the rate of increase slowing down.
\item Deltas w.r.t $L_{0}$ and $L_{1}$ are very small and close to 0 for
all swaptions, because they are ahead of all exercise time and are
mainly used for discounting.
\item With the increase of exercise times, magnitudes of Deltas w.r.t near
term Libor rates such as $L_{2}$ decrease, while magnitudes of Deltas
w.r.t farther term Libor rates such as $L_{9}$ increase. Here we
treat Delta in absolute value. The detail findings are listed below.
\begin{itemize}
\item For swptn\_1, Libor rates $L_{2}$, $L_{3}$, ..., $L_{9}$ are all
behind exercise time: $T_{2}$ and hence all of them are the underlyings
of swptn\_1. Therefore Deltas w.r.t $L_{2}$, $L_{3}$, ..., $L_{9}$
are very close.
\item For swptn\_2, Libor rate $L_{2}$ is the underlying of the first exercise
time: $T_{2}$, $L_{3}$, $L_{4}$, ..., $L_{9}$ are the underlyings
of the two exercise time: $T_{2}$ and $T_{3}$. Therefore compared
to swptn\_1, Delta w.r.t $L_{2}$ decreases, while Deltas w.r.t $L_{3}$,
$L_{4}$, ..., $L_{9}$ increase.
\item With the same logic, other remaining swaptions demonstrate similar
pattern. For example, for swptn\_9, $L_{2}$ is the underlying of
the first exercise time: $T_{2}$, $L_{3}$ is the underlying of the
first two exercise time: $T_{2}$ and $T_{3}$, ..., $L_{9}$ is the
underlying of all exercise time: $T_{2}$, $T_{3}$, ..., $T_{9}$.
Therefore we can see the clear pattern that $Delta(L_{2})<Delta(L_{3})<\cdots<Delta(L_{9})$
\end{itemize}
\item The computational speed is fast. We run the tests on one V100 GPU
and it takes around 1\textasciitilde{}2 minutes with quarterly time
step.
\end{itemize}
Therefore, we can conclude that the prices and Deltas from the backward
solver are effective and efficient. The backward solver is well-adapted
to price Bermudan swaptions. 

\begin{table}[H]
\makebox[\textwidth][c]{%
\begin{tabular}{|c|c|c|c|c|c|c|c|c|}
\hline 
 & {\footnotesize{}Swptn\_1} & {\footnotesize{}Swptn\_2} & {\footnotesize{}Swptn\_3} & {\footnotesize{}Swptn\_4} & {\footnotesize{}Swptn\_5} & {\footnotesize{}Swptn\_6} & {\footnotesize{}Swptn\_7} & {\footnotesize{}Swptn\_8}\tabularnewline
\hline 
\hline 
{\footnotesize{}Time (s)} & {\footnotesize{}64} & {\footnotesize{}79} & {\footnotesize{}95} & {\footnotesize{}106} & {\footnotesize{}127} & {\footnotesize{}135} & {\footnotesize{}137} & {\footnotesize{}149}\tabularnewline
\hline 
{\footnotesize{}NPV} & {\footnotesize{}0.003005} & {\footnotesize{}0.003553} & {\footnotesize{}0.003882} & {\footnotesize{}0.004097} & {\footnotesize{}0.004263} & {\footnotesize{}0.004412} & {\footnotesize{}0.004541} & {\footnotesize{}0.004628}\tabularnewline
\hline 
{\footnotesize{}Diff\_NPV} &  & {\footnotesize{}0.000548} & {\footnotesize{}0.000329} & {\footnotesize{}0.000215} & {\footnotesize{}0.000166} & {\footnotesize{}0.000149} & {\footnotesize{}0.000129} & {\footnotesize{}0.000087}\tabularnewline
\hline 
\end{tabular}}\caption{The NPVs of short term Bermudan swaptions and time needed for the
NPV calculations \label{tab:Sec4.4: The-NPVs-of-Bermudan-Swptn-1}}
\end{table}

\begin{table}[H]
\makebox[\textwidth][c]{%
\begin{tabular}{|c|c|c|c|c|c|c|c|c|}
\hline 
 & {\footnotesize{}Swptn\_1} & {\footnotesize{}Swptn\_2} & {\footnotesize{}Swptn\_3} & {\footnotesize{}Swptn\_4} & {\footnotesize{}Swptn\_5} & {\footnotesize{}Swptn\_6} & {\footnotesize{}Swptn\_7} & {\footnotesize{}Swptn\_8}\tabularnewline
\hline 
\hline 
{\footnotesize{}Time (s)} & {\footnotesize{}64} & {\footnotesize{}79} & {\footnotesize{}95} & {\footnotesize{}106} & {\footnotesize{}127} & {\footnotesize{}135} & {\footnotesize{}137} & {\footnotesize{}149}\tabularnewline
\hline 
{\footnotesize{}NPV} & {\footnotesize{}0.003005} & {\footnotesize{}0.003553} & {\footnotesize{}0.003882} & {\footnotesize{}0.004097} & {\footnotesize{}0.004263} & {\footnotesize{}0.004412} & {\footnotesize{}0.004541} & {\footnotesize{}0.004628}\tabularnewline
\hline 
{\footnotesize{}Diff\_NPV} &  & {\footnotesize{}0.000548} & {\footnotesize{}0.000329} & {\footnotesize{}0.000215} & {\footnotesize{}0.000166} & {\footnotesize{}0.000149} & {\footnotesize{}0.000129} & {\footnotesize{}0.000087}\tabularnewline
\hline 
\end{tabular}}\caption{The NPVs of short term Bermudan swaptions and time needed for the
NPV calculations \label{tab:Sec4.4: The-NPVs-of-Bermudan-Swptn}}
\end{table}

\begin{table}[H]
\makebox[\textwidth][c]{%
\begin{tabular}{|c|c|c|c|c|c|c|c|c|c|c|}
\hline 
{\footnotesize{}Swptn\textbackslash{}Underlying} & {\footnotesize{}$L_{0}$} & {\footnotesize{}$L_{1}$} & {\footnotesize{}$L_{2}$} & {\footnotesize{}$L_{3}$} & {\footnotesize{}$L_{4}$} & {\footnotesize{}$L_{5}$} & {\footnotesize{}$L_{6}$} & {\footnotesize{}$L_{7}$} & {\footnotesize{}$L_{8}$} & {\footnotesize{}$L_{9}$}\tabularnewline
\hline 
\hline 
{\footnotesize{}Swptn\_1} & {\footnotesize{}0.0119} & {\footnotesize{}-0.0077} & {\footnotesize{}-0.2445} & {\footnotesize{}-0.2443} & {\footnotesize{}-0.2410} & {\footnotesize{}-0.2440} & {\footnotesize{}-0.2388} & {\footnotesize{}-0.2367} & {\footnotesize{}-0.2362} & {\footnotesize{}-0.2369}\tabularnewline
\hline 
{\footnotesize{}Swptn\_2} & {\footnotesize{}0.0009} & {\footnotesize{}0.0055} & {\footnotesize{}-0.1634} & {\footnotesize{}-0.2469} & {\footnotesize{}-0.2413} & {\footnotesize{}-0.2421} & {\footnotesize{}-0.2387} & {\footnotesize{}-0.2374} & {\footnotesize{}-0.2373} & {\footnotesize{}-0.2392}\tabularnewline
\hline 
{\footnotesize{}Swptn\_3} & {\footnotesize{}-0.0010} & {\footnotesize{}-0.0024} & {\footnotesize{}-0.1394} & {\footnotesize{}-0.2013} & {\footnotesize{}-0.2453} & {\footnotesize{}-0.2475} & {\footnotesize{}-0.2406} & {\footnotesize{}-0.2420} & {\footnotesize{}-0.2408} & {\footnotesize{}-0.2440}\tabularnewline
\hline 
{\footnotesize{}Swptn\_4} & {\footnotesize{}-0.0146} & {\footnotesize{}0.0109} & {\footnotesize{}-0.1309} & {\footnotesize{}-0.1853} & {\footnotesize{}-0.2193} & {\footnotesize{}-0.2508} & {\footnotesize{}-0.2442} & {\footnotesize{}-0.2423} & {\footnotesize{}-0.2446} & {\footnotesize{}-0.2456}\tabularnewline
\hline 
{\footnotesize{}Swptn\_5} & {\footnotesize{}-0.0052} & {\footnotesize{}0.0133} & {\footnotesize{}-0.1260} & {\footnotesize{}-0.1776} & {\footnotesize{}-0.2106} & {\footnotesize{}-0.2343} & {\footnotesize{}-0.2455} & {\footnotesize{}-0.2446} & {\footnotesize{}-0.2429} & {\footnotesize{}-0.2470}\tabularnewline
\hline 
{\footnotesize{}Swptn\_6} & {\footnotesize{}-0.0157} & {\footnotesize{}0.0170} & {\footnotesize{}-0.1192} & {\footnotesize{}-0.1704} & {\footnotesize{}-0.2013} & {\footnotesize{}-0.2202} & {\footnotesize{}-0.2311} & {\footnotesize{}-0.2510} & {\footnotesize{}-0.2494} & {\footnotesize{}-0.2529}\tabularnewline
\hline 
{\footnotesize{}Swptn\_7} & {\footnotesize{}0.0125} & {\footnotesize{}0.0153} & {\footnotesize{}-0.1148} & {\footnotesize{}-0.1627} & {\footnotesize{}-0.1944} & {\footnotesize{}-0.2138} & {\footnotesize{}-0.2210} & {\footnotesize{}-0.2387} & {\footnotesize{}-0.2584} & {\footnotesize{}-0.2613}\tabularnewline
\hline 
{\footnotesize{}Swptn\_8} & {\footnotesize{}0.0063} & {\footnotesize{}0.0166} & {\footnotesize{}-0.1110} & {\footnotesize{}-0.1608} & {\footnotesize{}-0.1888} & {\footnotesize{}-0.2114} & {\footnotesize{}-0.2129} & {\footnotesize{}-0.2347} & {\footnotesize{}-0.2522} & {\footnotesize{}-0.2726}\tabularnewline
\hline 
\end{tabular}}\caption{The Deltas of short term Bermudan swaptions \label{tab:Sec4.4: The-Deltas-of-Bermudan-Swptn}}
\end{table}

\subsection{Long Maturity Bermudan Swaption Pricing using Backward Solver}

\begin{figure}[H]
\makebox[\textwidth][c]{\includegraphics[width=0.8\textwidth]{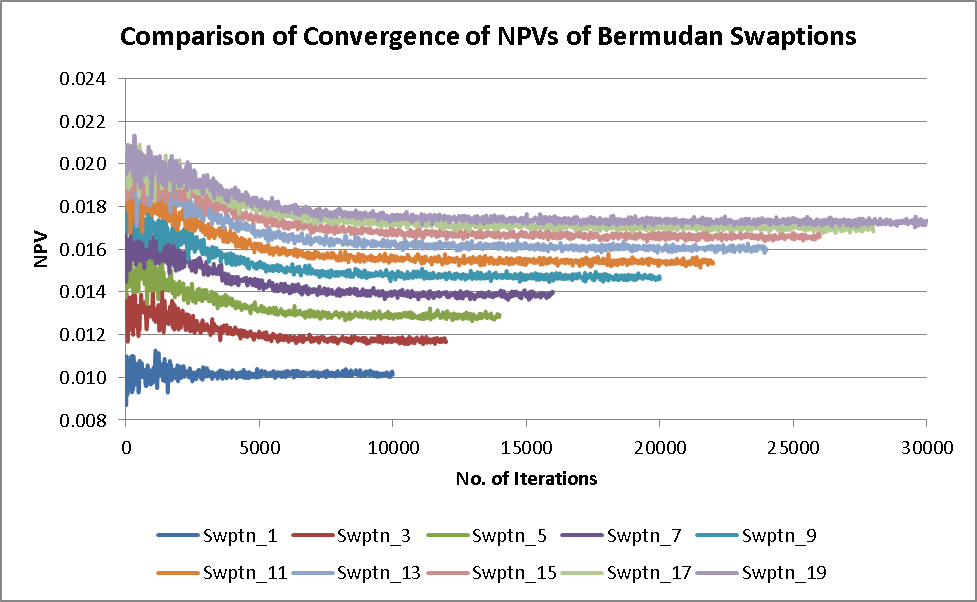}}

\caption{Convergence test of backward solver for NPVs of long term Bermudan
swaptions with different exercise times \label{fig:Sec4.5 Comparison-of-convergence-NPV-LTBermudan-Swptn}}
\end{figure}

\begin{figure}[H]
\makebox[\textwidth][c]{\includegraphics[width=0.8\textwidth]{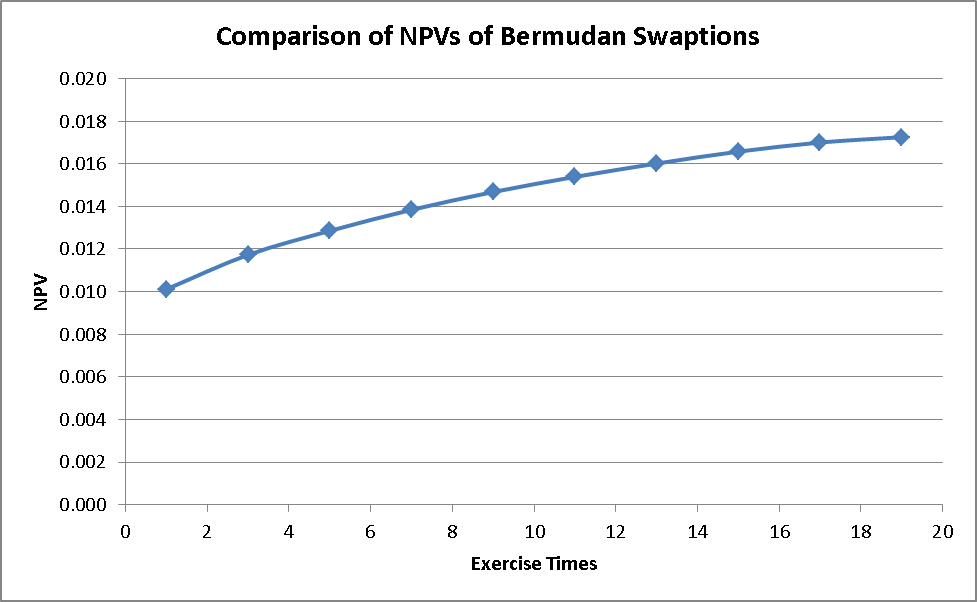}}

\caption{Comparison of NPVs for long term Bermudan swaptions with different
exercise times \label{fig:Sec4.5 Comparison-of-NPV-Diff-Bermudan-Swptn}}
\end{figure}

\begin{figure}[H]
\makebox[\textwidth][c]{\includegraphics[width=0.8\textwidth]{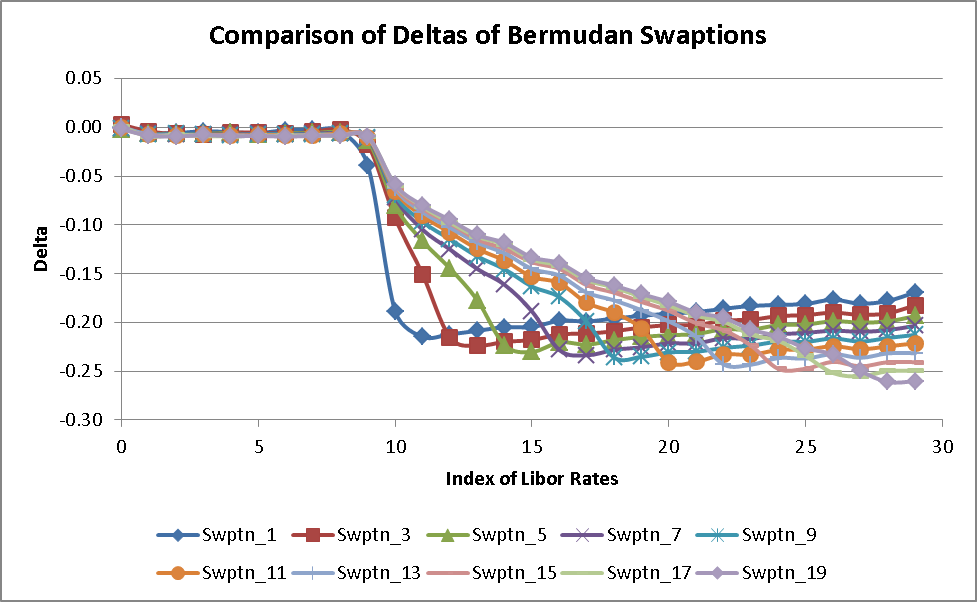} }

\caption{Comparison of Deltas for long term Bermudan swaptions with different
exercise times \label{fig:Sec4.5 Comparison-of-Deltas-LT-Bermudan-Swptn}}
\end{figure}

In this last example, we use backward solver to price long maturity
Bermudan swaptions. The Libor market model set-up, market information,
BSDEs and Neural Network setting are the same as those in Section
\ref{4.4-subsec:Short-Term-Bermudan}, except that the dimensions
of PDEs increase from 10 to 30 since there is a total of 30 Libor
rates involved. A total of 4096 simulated samples/paths are used for
optimization. The Bermudan swaptions we use in testing are listed
below.
\begin{itemize}
\item The exercise times of long maturity Bermudan swaption are listed in
the following table with each time interval around 0.5Yr. 
\begin{table}[H]
\makebox[\textwidth][c]{%
\begin{tabular}{|c|c|c|c|c|c|c|}
\hline 
{\footnotesize{}T} & {\footnotesize{}$T_{10}$} & {\footnotesize{}$T_{11}$} & {\footnotesize{}$T_{12}$} & {\footnotesize{}...} & {\footnotesize{}$T_{28}$} & {\footnotesize{}$T_{29}$}\tabularnewline
\hline 
\hline 
{\footnotesize{}Expiry} & {\footnotesize{}5.0667} & {\footnotesize{}5.5694} & {\footnotesize{}6.0861} & {\footnotesize{}\dots{}} & {\footnotesize{}14.2028} & {\footnotesize{}14.7083}\tabularnewline
\hline 
\end{tabular}}\caption{The exercise time of long term Bermudan swaption}
\end{table}
.
\item There are 10 Bermudan swaptions in total: Bermudan swaption 1 (swaption
1) contains one exercise time which is $T_{10}$; Bermudan swaption
3 (swaption 3) contains three exercise time which is $T_{10}$, $T_{11}$,
$T_{12}$; ...; Bermudan swaption 19 (swaption 19) contains 19 exercise
time which is $T_{10}$, $T_{11}$, $T_{12}$, ..., $T_{28}$.
\item Their underlying swaps have the same terminal time, which is $T_{30}=15.2194Yr$. 
\item They have the same notional, which is 1 and they have the same strike,
which is 2.7155\%.
\end{itemize}
Table \ref{tab:Sec4.5: The-NPVs-of-Bermudan-Swptn} and Figure \ref{fig:Sec4.5 Comparison-of-convergence-NPV-LTBermudan-Swptn}
through Figure \ref{fig:Sec4.5 Comparison-of-Deltas-LT-Bermudan-Swptn}
show the NPVs and Deltas of long term Bermudan swaptions. From the
results, we have the following observations:
\begin{itemize}
\item The convergence of NPVs is steady. With increasing number of exercise
times, the convergence speed slows down. Therefore we use more iterations
for Bermudan swaptions with more exercise times.
\item With the increase of exercise times, NPVs also increase quite smoothly
with the rate of increase slowing down.
\item Deltas w.r.t the first 10 Libor rates are very small and close to
0 for all swaptions, because they are ahead of all exercise time and
are mainly used for discounting.
\item With the increase of exercise times, magnitudes of Deltas w.r.t near
term Libor rates such as $L_{10}$ decrease, while magnitudes of Deltas
w.r.t further term Libor rates such as $L_{29}$ increase. Here we
treat Delta in absolute value. The detail findings are as follows:
\begin{itemize}
\item For swptn\_1, Libor rates $L_{10}$, $L_{11}$, ..., $L_{29}$ are
all behind exercise time $T_{10}$ and hence all of them are the underlyings
of swptn\_1. Therefore Deltas w.r.t $L_{10}$, $L_{11}$, ..., $L_{29}$
are close with the trend going down due to the discounting.
\item For swptn\_2, Libor rate $L_{10}$ is the underlying of the first
exercise time $T_{10}$, $L_{11}$, $L_{12}$, ..., $L_{29}$ are
the underlyings of the two exercise time $T_{10}$ and $T_{11}$.
Therefore compared to swptn\_1, Delta w.r.t $L_{10}$ decreases, while
Deltas w.r.t $L_{11}$, $L_{12}$, ..., $L_{29}$ increase.
\item With the same logic, other remaining swaptions have similar pattern.
For example, for swptn\_19, $L_{10}$ is the underlying of the first
exercise time $T_{10}$, $L_{11}$ is the underlying of the first
two exercise time $T_{10}$ and $T_{11}$, ..., $L_{28}$ is the underlying
of all exercise time $T_{10}$, $T_{11}$, ..., $T_{28}$. Therefore
we can see the clear pattern that $Delta(L_{10})<Delta(L_{11})<\cdots<Delta(L_{28})$
\end{itemize}
\item The computational speed is fast. We run the tests on one V100 GPU
and it takes about 3 minutes for swptn\_1, 4.5 minutes for swptn\_3,
... , 20 minutes for swptn\_19. The time step is quarterly.
\end{itemize}
Therefore, we can conclude that the prices and Deltas from the backward
solver are reasonable. The backward solver is effective to price the
Bermudan swaption with long maturity. 

\begin{table}[H]
\makebox[\textwidth][c]{%
\begin{tabular}{|c|c|c|c|c|c|c|c|c|c|c|}
\hline 
 & {\footnotesize{}Swptn\_1} & {\footnotesize{}Swptn\_3} & {\footnotesize{}Swptn\_5} & {\footnotesize{}Swptn\_7} & {\footnotesize{}Swptn\_9} & {\footnotesize{}Swptn\_11} & {\footnotesize{}Swptn\_13} & {\footnotesize{}Swptn\_15} & {\footnotesize{}Swptn\_17} & {\footnotesize{}Swptn\_19}\tabularnewline
\hline 
\hline 
{\footnotesize{}Time(Min)} & {\footnotesize{}3} & {\footnotesize{}4.5} & {\footnotesize{}6} & {\footnotesize{}7.5} & {\footnotesize{}10} & {\footnotesize{}12} & {\footnotesize{}14} & {\footnotesize{}16} & {\footnotesize{}18} & {\footnotesize{}20}\tabularnewline
\hline 
{\footnotesize{}NPV} & {\footnotesize{}0.01012} & {\footnotesize{}0.01173} & {\footnotesize{}0.01286} & {\footnotesize{}0.01384} & {\footnotesize{}0.01468} & {\footnotesize{}0.01539} & {\footnotesize{}0.01601} & {\footnotesize{}0.01657} & {\footnotesize{}0.01700} & {\footnotesize{}0.01725}\tabularnewline
\hline 
{\footnotesize{}Diff\_NPV} &  & {\footnotesize{}0.00161} & {\footnotesize{}0.00113} & {\footnotesize{}0.00099} & {\footnotesize{}0.00084} & {\footnotesize{}0.00071} & {\footnotesize{}0.00062} & {\footnotesize{}0.00056} & {\footnotesize{}0.00043} & {\footnotesize{}0.00025}\tabularnewline
\hline 
\end{tabular}}\caption{The NPVs of long term Bermudan swaption and time needed for the NPV
calculations \label{tab:Sec4.5: The-NPVs-of-Bermudan-Swptn}}
\end{table}

It is quite encouraging that the DNN-based BSDE backward solver works
well for Bermudan swaptions of all maturities.

\section{Conclusion\label{sec:Conclusion}}

In this work, we have developed a deep neural network-based BSDE solver
for Libor Market Model (LMM) based on the recent works on Artificial
Intelligence and Machine Learning developments. Our forward deep learning
implementation adopts a similar approach as the one explored by Weinan
E et al. \cite{Weinan17,Jiequn17}, which uses neural network to solve
high dimensional parabolic PDEs (or their equivalent BSDEs). However,
the standard forward DNN solver is only suitable for pricing European
style options. Therefore, in order to price callable derivatives such
as Bermudan interest rate options, we developed a backward DNN solver,
which has a parallel neural network structure as forward solver. Our
new methodology projects the option price backward to the initial
time with the loss function given by the variance of the projected
option price at initial time. We have tested a few numerical examples
to illustrate that the backward solver works as accurately as forward
solver and Monte Carlo simulation method in QuantLib. The prices of
Bermudan swaption calculated using backward DNN solver are quite reasonable.
These deep neural network based solvers have clear advantage over
traditional methods: 

(1) They can produce gradients of option price (Delta) directly. 

(2) They are highly scalable and can be applied to many high dimensional
PDEs without changing its implementation. 

(3) They can easily be implemented in Python using TensorFlow, which
runs efficiently in GPU. 

(4) The capability of backward solver to price Bermudan swaption makes
deep learning-based solver suitable for a wide range of different
applications, since most of the traded financial instruments in hedging
risks are Bermudan style rate options such as Bermudan style swaptions,
Callable CMS spread options, and Callable Range Accruals, etc. 

This general methodology can also be applied to other asset groups
such as equity, foreign exchange, commodity, and various value adjustment
processes. Our immediate further work will be focused on putting more
structures in LMM with DNN implementation. Some theoretical questions
in mathematics and DNN structure are interesting for us to explore
when we will apply this new approach in our model risk management,
including model validation. 

\newpage{}

\end{document}